\documentclass[%
final,
twocolumn,
 amsmath,amssymb,
 aps,
pra,
floatfix,
]{revtex4-2}

\usepackage{enumerate}
\usepackage{graphicx}
\usepackage{dcolumn}
\usepackage{bm}
\usepackage{url, esint}
\usepackage{multirow}
\usepackage{physics,mhchem}
\usepackage[ 
  breaklinks,
  colorlinks=true,
  bookmarks=true,
  bookmarksopenlevel=3,
  bookmarksnumbered=true,
  menucolor=blue,
  linkcolor=blue,
  citecolor=blue,
  urlcolor=blue,
  pdfkeywords={},
  pdfstartview=FitH]{hyperref}

\def\CO2{\ce{CO2}}
\def\O2{\ce{O2}}
\def\N2{\ce{N2}}

\begin{document}

\preprint{arXiv:2022}

\title{A course on Climate Change and Sustainable Building Design}

\author{Claire Akiko Marrache-Kikuchi}
\email{claire.marrache@universite-paris-saclay.fr}
\affiliation{Université Paris-Saclay, CNRS, IJCLab, 91405, Orsay, France.}
 
\author{Guillaume Roux}%
\email{guillaume.roux@universite-paris-saclay.fr}
\affiliation{%
Université Paris-Saclay, CNRS, LPTMS, 91405, Orsay, France.
}%

\author{Jean-Marie Fischbach}%
\affiliation{%
Magistère de Physique Fondamentale, Université Paris-Saclay, 91405, Orsay, France.
}%

\author{Bertrand Pilette}%
\affiliation{%
Magistère de Physique Fondamentale, Université Paris-Saclay, 91405, Orsay, France.
}%

\date{\today}

\begin{abstract}
We present an intermediate-level course on sustainable physics, which combines lectures and student projects. 
Sustainable physics concepts are progressively introduced through both a global and a specialized perspective: climate change and building design.
The lectures and hands-on activities on both topics show how they share common concepts and provide complementary points of view.
Climate change mitigation and adaptation are touched on via student group projects, where they are asked to tackle a specific question of their choosing via reviews of the literature, modeling or experiments. 
\end{abstract}

\maketitle

\section{\label{sec:intro}Introduction}

Climate change, and environmental issues in general, are a growing source of concern among the younger generation, especially in high-income countries. \cite{nepravs2022climate, lawson2019children} Like other students, those studying physics show a strong appetite for courses that tackle these questions. 
Teaching about climate change is challenging because there are more problems than solutions, which generates anxiety among students. \cite{ojala2021anxiety} 
Additionally, many students get information on these topics from videos or popular articles.  These passive communication channels tend to leave students feeling helpless.

The community of researchers and educators also displays a growing concern regarding climate change. In particular, the physics community feels that it has the scientific background to tackle these problems, though it usually recognizes that an interdisciplinary approach is critically needed. 
Instructors are often hesitant to teach topics that are outside of their areas of expertise. 
Moreover, undergraduate-level courses need to provide a bridge between high-school-level and graduate-level education on energy and climate change.
Because physicists should, in principle, be able to teach all courses at the bachelor's level, we feel that there is a strong motivation for developing general environmental physics materials for and by non-experts but with a university-level science content.

There are several ways to approach systemic issues. One is the global approach, by discussing data and physical mechanisms of global impact with regard to energy and climate. Another approach is to start from a focused topic, as an example, and to progressively draw the threads towards all the global issues to which it is connected.

In this paper, we describe an intermediate-level course for physics students titled 
"Physics for sustainable development", which tries to address these challenges in two main ways.
First, the structure of the course encourages active learning. The lectures integrate hands-on tutorials and flipped classrooms; and the second half of the course reserves time for group projects. We believe that this active approach to learning dispels anxiety.
Second, we combine teaching about global climate change problems with explorations of a specific local solution: sustainable building construction.
This choice came from the expertise of one of us and makes many relevant connections with global issues. Architecture and housing also provide compelling topics for projects.
One of the pleasures and motivations for us is to bring scientific reasoning and physicists' way of thinking outside the typical physics realm. We show students that physics provides them with skills to approach both complex societal issues and everyday life situations. 


Before describing the contents and the philosophy of this course in more details in the next sections, let us mention the context in which it was developed. 
The course was offered as an elective to third year physics students at Paris-Saclay University and started in spring 2020. The course typically enrolled around 15 motivated students.
We begin with 12 hours of class sessions: six hours on climate change and six hours on building design-related problems. The second part of the course is dedicated to group projects, covering six two hour-long sessions. Students are expected to spend about the same amount of time working outside class on the lectures and the project.


\section{\label{sec:cours}Class sessions}

\begin{table*}[t!]
\begin{ruledtabular}
\begin{tabular}{p{0.05\linewidth}p{0.095\linewidth}p{0.34\linewidth}p{0.085\linewidth}p{0.39\linewidth}}
Week & Discussed in section & Topic & Duration & Main covered topics \\
\hline
1 & \ref{subsubsection_Anthropocene} & Anthropocene & 1 hour & Ozone depletion, great acceleration, biogeochemical cycles\\
1 \& 3 & \ref{subsubsection_Climate} &  Climate change & 2 hours & Radiative balance, greenhouse effect, energy redistribution, anthropic perturbations, modeling and projections, probes and impacts of global warming\\
2 & \ref{subsubsection_Urbanization} & Urbanization - What is at stake? & 1 hour & World urban demography, critical review of some sustainable building practices\\
2 & \ref{subsubsection_Bioclimatic} & Principles of bioclimatic construction & 1 hour & Assessing the climate of a building site, influence of the building conception on its sustainability\\
3 \& 5 & \ref{subsubsec:Energy_GR} & Energy and society & 2 hours & Energy supply chain, everyday life orders of magnitudes, energy mix, fossils energies, connection to economy and development \\
4 & \ref{subsubsection_Energy_bldgs} & Energy in buildings & 2 hours & Review of energy sources for a building, energy production and recovery systems\\
5 & \ref{subsubsection_Challenges} & Challenges for sustainable development & 1 hour & Geography with orders of magnitudes on population and territories, global views on sustainable development challenges \\
6 & \ref{subsubsection_Thermal_bldgs_city} & Thermal considerations in buildings and in the city & 2 hours & Heat transfer mechanisms, urban Heat Island Effect\\

\end{tabular}
\end{ruledtabular}
\caption{\label{tab:table1}%
A possible chronological sequence of class sessions alternating between topics of sections II.B and II.C. Other sequences could be imagined.}
\end{table*}

The class sessions are deliberately designed 
to provide a broad overview of the problems that cause and result from climate change (see table \ref{tab:table1}). We do not shy away from discussing topics outside physics such as sociology, economics, politics or behavioral sciences. Our approach tries to treat climate change as a complex system and to highlight the link between the various fields. The ultimate goal is to have students question their actions as well as the society they live in, and develop critical thinking. Moreover, climate change is not our research specialty; it is neither our place nor our wish to give a specialized lecture on a given subject. 

Sessions on climate change by Guillaume Roux and on building design by Claire A. Marrache-Kikuchi alternate, and, whenever possible, call on each other to create a coherent syllabus.
More precisely, the course is divided into units (topics), each of which consists in a two-hour session that includes both a lecture and problem-solving tutorials. Each topic is described in the corresponding subsection below. 
We emphasize that estimates of orders of magnitude and elementary modeling through Fermi questions are crucial skills in the education of physicists. Our first learning objective for students is to be able to handle these problems. This is all the more relevant for tackling real-life physics problems and complex systems, such as climate, as we do in this course. Examples that we mention in class are detailed in the supplementary material.\cite{SupplMat}

The lecture part of the course is evaluated by a two hour-long written exam comprising multiple choice questions and tutorial-like exercises with Fermi questions or elementary modeling.
Examples of problems (and their answers) are given in Section IV of the supplementary material.\cite{SupplMat} 
The students' scores at the exam are varied. Out of the 40 students who have enrolled in the course since its creation three years ago, 89\% of students passed the exam. The average grade was 13.3/20 (standard deviation 2.6). The dispersion in the results probably indicates the difficulty most students have with elementary modeling and Fermi questions. We believe that this is to be expected, since most students never had to do this kind of reasoning before the course.

Regarding the scientific level of the lectures, we go into more detail than we would for first year students, but do not specialize to the point of graduate level courses.
Let us note that, at Paris-Saclay University, most of our students have already had a general introduction about environmental challenges during their second year, which is mandatory for all bachelor students, whatever their major. It is a Small Private Online Course (a MOOC that is only accessible to students of our university), the contents of which are freely available in Ref.~\cite{lourtioz2021enjeux} (in French). Although this lecture-based course presents several physical concepts, the level of the physics is low.

In our course, the lectures are quite dense, but are usually useful to give students an overview of the diversity of issues and fill the gaps in their knowledge. Out of the 40 students who have attended the course, we have gathered the feedback of 31 of them. 
Their evaluation of the lecture part of the course is satisfactory, though with a significant variability: the average rating for this part on a 5-point scale was 3.7 (standard deviation 1.0).

\subsection{\label{subsec:basis}Common concepts}

Before entering into the details of the different subjects, we would like to highlight that, although they seem unrelated,
climate and building design share common concepts emanating from physics, but not only. 
In this section, we list some of these concepts, since they can be taught in relation either with climate change or building design, and reused in the other topic, with the satisfaction of observing each time the universality of physics.

A first subject is radiation, including Planck's and Stefan-Boltzmann's laws. In the \textit{Solar constant} activity \cite{SupplMat}, we use the Stefan-Boltzmann law to determine the average power a given location receives from the Sun during daylight, called the solar constant $G_{SC}\simeq 1360$ W.m$^{-2}$. The concepts of emissivity, absorptivity, reflectivity, transmissivity and radiation heat transfer mechanism all enter into play both in the Earth radiative balance and in heat transfers in buildings. 
Convection, conduction and diffusion are naturally relevant too. Among others, Ref.~\cite{HeatTransfer} offers a clear presentation of these topics, full of everyday examples.

More generally, whenever possible, we try to establish a connection between the theoretical concepts (e.g. thermodynamics) seen in basic physics courses, and their consequences for practical systems (e.g. thermal machines, heat losses, evaporative energy\ldots). We find that this does not necessarily require a sophisticated mathematical treatment, and enables the description of a wide variety of phenomena and practical realizations. Fluid mechanics is another common tool with applications ranging from the atmospheric circulation to natural ventilation in buildings or wind turbines.



Beyond physics, basic concepts of chemistry and biology are useful.  General concepts about climate classification and biomes are actually determinant for urbanization and civilization development. As a common starting point, we ask the class to work in groups to determine the climate in a site of their choosing and to discuss its impact on building conception (see the \textit{Climate assessment} activity \cite{SupplMat}).\cite{climate_activity} 
Similarly, trees play a role in biogeochemical cycles but also in urban planning. In the \textit{Cooling effect of trees} activity \cite{SupplMat}, we estimate the energy necessary to evaporate the 450 L of water produced daily by a large tree, and find that it is equivalent to that of 6 domestic coolers ($\sim$ 2.5 kW) running for 19 hours per day.


\subsection{\label{subsec:climate}Climate change}

Here we present the structure of the climate portion of the course. The subject being very broad, we decided to emphasize two main goals. The first is to contextualize the issue in order to gain a broader perspective. 
The second goal is to give the salient scientific facts explaining the human-caused climate change. 
In passing, we define and explain some essential concepts that appear in Intergovernmental Panel on Climate Change (IPCC) reports.\cite{IPCC} To this end, we aim at making the students comfortable reading more advanced literature such as IPCC reports and research papers. These lectures are centered on climate change but actually go beyond the physics of the climate system by tackling typical environmental physics topics for which a useful Resource Letter~\cite{Forinash2019} is available.

\subsubsection{Anthropocene}
\label{subsubsection_Anthropocene}

The course opens with a one-hour lecture on the ``Anthropocene". 
The idea is to embed climate change among the global impacts of humans and discuss key numbers of the ``great acceleration".\cite{steffen2015,fressoz2016evenement,ellis2018anthropocene}
Through the example of ozone depletion and the perturbation of global biogeochemical cycles, we understand 
the interconnection of the main challenges of environmental science.
One motivation is that physicists often have a techno-solutionist bias of climate change mitigation, that consists of trying to``restore" the carbon cycle. However, a global perspective shows that carbon-focused solutions most often raise issues in related environmental domains, for instance land use, water use, nitrogen and phosphorous cycles, or biodiversity.\cite{dolman2019biogeochemical}
Yet the great advantage of the carbon cycle is that it is very well documented and allows for many ``Fermi questions" which keep students focused. Doing the quantitative calculations by oneself raises awareness. In example II.A in the supplementary material~\cite{SupplMat}, we give the example of a class activity estimating how many gigatons of carbon have to be emitted to increase the atmosphere concentration by 1 ppm.


\subsubsection{Climate change}
\label{subsubsection_Climate}

This lecture proposes an overview of the main concepts of climate science with few mathematical aspects. The focus is on understanding the orders of magnitude, the time scales, and the main components and mechanisms of the climate system.
The emphasis is placed on the radiative balance and greenhouse effect for its importance in current global warming.
Another focus is on climate observed impacts and climate projections, with or without human-caused sources.
We define radiative forcing, feedback, climate sensitivity and transient response. 
The scientific level, spirit, arguments, and contents that are developed are strongly inspired by three books: \cite{melieres2015climate}, \cite{archer2011global} and \cite{Delmas2007}.
\textit{American Journal of Physics} also provides two recent Resource Letters on climate change.\cite{RessourceLetterClimate1, RessourceLetterClimate2}
Additional and updated figures are found in specialized resources, in particular the IPCC reports.
The question of the impact of climate change on societies opens the discussion beyond scientific facts.
For an example of class activity on the modeling of radiative balance and understanding the greenhouse effect, see II.B in the supplementary material~\cite{SupplMat}.

\subsubsection{Energy and society}
\label{subsubsec:Energy_GR}

This topic naturally follows in a course for physicists, since fossil energies represents about 75\% of greenhouse gases emissions. After recalling some basics of energy and power, energy supply chains are presented in order to help students understand the figures and definitions used in the economics of energy.
Scenarios about the possible evolution of global and national energy mix are presented.
The physics of the main energy sources is discussed taking a broad perspective and using students background.
Dedicated to everyday and industrial energy uses, this topic allows for many Fermi questions and estimates for energy consumption or power requirements.
Here, the spirit is certainly inspired by the celebrated book by David Mackay \cite{mackay2009sustainable}, freely available, and also the recent well written textbook by Jaffe and Taylor~\cite{jaffe_taylor_2018}.
Another source of inspiration for this topic is the French popularizer Jean-Marc Jancovici~\cite{Janco}, who stresses society dependencies upon fossil energies, with similar viewpoints as Vaclav Smil.\cite{smil2018energy}

Although this lecture deals with the most physics-related topics, most ideas are new for students, despite the fact that they usually have some qualitative knowledge about energy supply chains. The lecture is founded on usage in terms of quantity, but includes discussions of lifestyle. This is a good place to introduce concepts such as the Jevons paradox~\cite{ALCOTT20059}, technological lock-in~\cite{UNRUH2000817} and the coupling between economic growth and energy consumption.\cite{kallis2018degrowth} The ideas are conveyed through examples. We find the example of cars, 
from their energy supply chain to their social impacts and the corresponding reshaping of urbanization, enlightening.

This class session is also well suited for discussing the finite resources problem in environmental science. The concept of peak oil is well known in the field of fossil energy but supplies of materials which are critical for renewable energy such as copper, rare-earth elements, cobalt, and lithium may also limit development.~\cite{valero2021material}
A historical perspective eventually opens a discussion at the crossroads between physics and social sciences.
For an example of class activity related to energy in everyday life, see II.C in the supplementary material~\cite{SupplMat} for the example of the power and CO$_2$ emissions of thermal cars. Other examples are food and batteries for energy storage.

\subsubsection{Challenges for sustainable development}
\label{subsubsection_Challenges}

In the final unit, we talk about subjects that do not usually belong to physics, which lead to controversial questions about human development.  Our motivation is to show students both the power and the limitations of technical and scientific perspectives in relation to global issues. We discuss figures about land occupation on Earth, population densities, and other basics geographical facts. We teach some aspects of global environmental history~\cite{testot2017cataclysmes}, showing how the industrial revolution led to the 20$^{\text{th}}$ century's great acceleration.  We discuss demographic challenges~\cite{lundquist2015demography,harper2018demography,veron2013demographie} as well as other limitations of human development.  We introduce the debate over historical responsibilities for global climate change and its inequitable impacts. We do not attempt to answer these questions, but we show students how they are addressed by the social sciences community.  For an example of a class activity on demography, in which modeling population growth helps to understand the social roots of the main demographic indicators, see II.D. in the supplementary material.~\cite{SupplMat}

\subsection{\label{subsec:arhi}Sustainable Building Design}
Architecture is a key industry in relation to climate change. The United Nations estimate that, in 2020, 55\% of the world population lived in urban areas and predicts that by 2050 this figure will be close to 70\%.\cite{urban_pop} This means that about 2.5 billion new urban dwellers will arrive in fast-expanding cities mainly located in South-East Asia, India, Africa, and Central and South America.\cite{city_growth} These new urban centers will need infrastructures, office buildings, and dwellings, the construction of which represents a huge challenge in a short time frame, and given the commitments of the 2016 Paris Climate Agreement.

In 2020, the building industry - including building operations, building materials manufacturing, and construction - represented 35\% of the world's energy consumption, and almost 40\% of the global CO$_2$ emissions.\cite{bldg_emissions} This industry is the second largest producer of primary plastic \cite{plastic}, and the main source of material consumption \cite{Materials2020}, with consequences that may be catastrophic both socially and environmentally. For instance, the mining of the sand needed to fabricate concrete is responsible for land loss as well as for the development of criminal groups.\cite{sand1, sand2} 

Concomitantly, architects explore new ways of building, promote the use of new materials and technologies, and revisit traditional building techniques, to make new constructions more sustainable. Architecture and the building industry as a whole is therefore experiencing an upsurge of ideas and innovation, which can in turn be appraised by scientists.

From a pedagogical point of view, the topic of building design in the context of climate change give us the opportunity to focus on a topic that is familiar to students. The systems are smaller - a house, a building, a neighborhood - and the stakes seem more manageable than when tackling climate change as a whole. These are also ideal systems to put various concepts of physics into practice.

We chose a flipped learning approach: students are required to have read a chapter on the topic at hand before class. At the beginning of the class, some examples are discussed, but most of the time is spent in solving Fermi questions, or in group activities. For each topic, we will detail below both the content of the written lecture notes \cite{SupplMat} and the class activities.

\subsubsection{Urbanization - What is at stake?}
\label{subsubsection_Urbanization}
This part of the course starts with an hour-long introductory lecture, the only one in this part that is not a flipped classroom. After presenting a few figures on the demography and geography of the world urban population, the impact of the building industry is assessed in terms of energy and material consumption. A few issues are stated, and some of the proposed solutions critically discussed to highlight the fact that none is ever perfect, but also to debunk some unreasonable performance promises or green-washing advertising.

One example is the use of recycled shipping containers to build either temporary \cite{container1} or cheap permanent housing.\cite{container2} This construction method uses recycled materials, reduces material waste, and can be completed rapidly. However, those who advocate retrofitting these containers into dwellings rarely considers the thermal performance of this envelope, and the environmental cost of heating, cooling, and ventilating these structures can easily exceed the benefit of their construction method. 

Commercial air filtering solutions to reduce particulate matter pollution are very diverse, and include urban infrastructures that vacuum air particles through a high voltage (30 kV) ionizing chamber \cite{smog_tower}, moss modules that claim to have the same air-cleaning capability as more than 250 trees \cite{city_tree}, and TiO$_2$-based photocatalytic fabrics that remove NO$_x$ compounds from the air to produce nitric acid.\cite{air_fabric} We discuss whether the advertised performance of these air-filtering techniques are realistic, and what the drawbacks can be in terms of public safety, energy consumption, or toxic release.  

In the same spirit, the advantages and drawbacks of photovoltaic energy, wind turbines, hydropower, geothermal energy, or nuclear energy are discussed. Solutions for water management \cite{NWRM} can be debated in light of the urban or rural context. The sustainability and economics of urban farms are also potential topics for discussion and debate. \cite{goldstein2016urban}

\subsubsection{Principles of bioclimatic construction}
\label{subsubsection_Bioclimatic}
This topic aims at understanding the physical principles underlying bioclimatic construction \cite{brophy2012green}. Students are also expected to exercise critical thinking to assess the sustainability of purpoted "green" or "sustainable" projects.

The first step in designing a sustainable building is to characterize the climate of the site. In this course, this is the pretext to determine the energy that is available from solar radiation (see the \textit{Solar constant activity} \cite{SupplMat}), introduce the use of a heliodon for measuring the sun's local angle of incidence, and reflect on the influence of wind speed and rainfall on a building (see the \textit{Climate assessment} activity \cite{SupplMat, comment_climate_assessment_activ}). The influence of the building environment, such as urbanization and terrain, is also examined. The wind turbulence in urban environments can, for example, be studied as a practical application of elementary fluid mechanics. 

We then continue by studying the influence of building design on energy consumption. The surface-to-volume ratio of a building is, for instance, an important parameter: the higher it is, the higher the energy demand, the higher the cooling effect of the wind, the more the soil is impervious to water, but the larger the natural lighting, and the larger the area of photovoltaic modules that can be installed. We consider how the projected use of the building affects the optimal window size and building orientation with regard to heating and lighting demands. \cite{paule2015current} We quantitatively evaluate the importance of albedo for external surfaces (façades and roofs) through the case study of the "Cool Roofs" initiative in New York \cite{gaffin2012bright}: knowing the global irradiation in New York, one can estimate the surface temperature of a layer of black or white gravel and show that the difference in albedo is responsible for a surface temperature difference of several tens of degrees (see the detailed derivation in the supplementary material \cite{SupplMat}). The basic definitions used for lighting are given (luminance, flux, emittance, intensity), and lighting solutions are discussed, including circadian lighting using LEDs and the transport of natural daylight in a building through optical fibers or reflecting pipes. Water collection devices, water recycling capabilities, and water purification systems are introduced.

This lecture ends by listing the most commonly used sustainability certification labels. A possible activity or homework consists in reviewing the criteria a particular certification label requires  (for new or retrofitted buildings, infrastructures or neighborhoods) (see the \textit{Assessing the sustainability labels} activity \cite{SupplMat}). Students can then make up their own mind on how stringent or effective the certifications are.

The embodied energy of buildings \cite{chastas2016embodied, cabeza2013low, dixit2010identification} and the life cycle analysis of building materials \cite{sharma2011life, ramesh2010life} could be interesting additional topics to discuss. Due to time constraints, we have chosen to leave these subjects for the interested students to tackle as projects. 

As a homework assignment, we ask students to investigate one physical quantity (solar irradiation, energy, water consumption, etc) and to establish a few orders of magnitude for these quantities (see the \textit{Orders of magnitude} activity \cite{SupplMat}). This exercise is designed to make them manipulate orders of magnitude and connect physical quantities to their everyday life.

\subsubsection{Energy in buildings}
\label{subsubsection_Energy_bldgs}
This topic closely follows the lecture on energy in relation to climate change (section \ref{subsubsec:Energy_GR}). First, we list the energy demands in buildings: heating, cooling, dehumidification, hot water production, air handling units, fan coil units, and the various electrical appliances the importance of which may vary depending on the building's usage (dwelling, office, factory). The coefficient of performance (CoP) for energy production devices is defined. Various sources of energy are then discussed \cite{Energy_sources}: geothermal energy, solar thermal collectors, photovoltaic panels (see the \textit{Determining a site's photovoltaic potential} activity \cite{SupplMat}), and wind turbines (see the \textit{Wind Power} activity \cite{SupplMat}). Energy production systems such as boilers, heat pumps, cogenerators, chillers and cooling towers, as well as energy recovery systems such as dual flow ventilation systems and energy recovery from gray water are mentioned. The aim of this lecture is for students to have a very broad overview of the various energy systems that are used in the construction industry, and to discuss their advantages and downsides.

When mentioning geothermal energy, we also study two examples in depth. The first one deals with underground buildings. Throughout the world, there are examples of underground buildings, both modern and indigenous.\cite{Underground_bldg} These buildings are particularly interesting examples for physicists because they make use of the thermal properties of the ground, based on the fact that daily or seasonal thermal fluctuations at the Earth's surface are strongly damped - by almost of factor of 10 - at a depth of 5 m.\cite{song2014effectiveness} Moreover, through its large thermal inertia, the soil acts on the corresponding building's facades as a thermal bath.

The second example we study in class is the principle of thermal labyrinths \cite{song2014effectiveness} and Canadian wells (see the \textit{Estimation of a Canadian well's efficiency} activity \cite{SupplMat}). These are very simple devices that draw the fresh air needed to ventilate a house or a building through an underground concrete labyrinth or a buried plastic pipe. The incoming air is therefore warmed in winter and cooled in summer, taking advantage of the ground's quasi-constant temperature throughout the year. The activity is designed to introduce modeling methodology. The objective is not to obtain an exact solution, but rather to outline how a physicist would reason to have an estimate of the answer (in the present case: by how much can you increase or decrease the temperature of the incoming air by having it go through a Canadian well). The modeling methodology (start with the simplest possible system and make simplifying assumptions before moving on to a more complex description) is a soft skill that will be useful to students whatever their future careers. In the realm of architecture, engineering consultants often keep all the building's complexity when simulating its thermal response for instance via dedicated software. The process is long and relatively computation-intensive, so that their advice cannot be readily actionable in a building's design phase. A simplified simulation, \textit{à la physicist}, by contrast, may give in a few hours an approximate solution that may be used by architects to optimize their design.


\subsubsection{Thermal considerations in buildings and in the city}
\label{subsubsection_Thermal_bldgs_city}
The last lecture deals with heat transfer mechanisms, which are one of the main issues for the construction industry in relation to climate change. Although the thermal performance of new dwellings and offices has improved in the past decades, at least in richer countries, heating and cooling still account for about 50\% of the energy needs in a building \cite{IEA_buildings}, and for about 30\% of the global CO$_2$ emissions.\cite{bldg_emissions} At larger scales, the urban heat island effect \cite{sobstyl2018role, fitria2019impact} has rendered cities both less comfortable to live in and more in need of cooling solutions. 

First, the basics of heat transfer phenomena are recalled: natural and forced convection, radiation and conduction. The notion of thermal resistance is introduced on an example taken from the building sector. For instance, one can calculate the effective thermal resistance of a window pane composed of a glazing fitted within a wooden frame, very much like one would determine the effective resistance of a resistor network. Determining the temperature in an igloo (see the \textit{Temperature in an igloo} activity \cite{SupplMat}) is a nice example of the competition between radiative and conductive heat transfers \cite{gonzalez2001intriguing}. This activity is also designed to encourage students to make assumptions, for instance on the igloo size or its occupation. We expect them to come up with an order of magnitude of the temperature, show them how this result depends (or more likely scarcely depends) on the details of the model, and compare it to measurements that are reported online.\cite{igloo1,igloo2}

After an assessment of all possible heat flows that come into play in a building, we also introduce the notion of thermal inertia. When considering the electrical analogy of the heat equation, thermal inertia acts as a capacitor $C$, so that a building can be modeled by a (complex) $RC$ circuit, where the electrical resistance $R$ is analogous to the thermal resistance in the initial problem. This is actually at the basis of an energy performance measurement method for houses.\cite{mangematin2012quick} Evaporative cooling is also mentioned: latent heat associated with phase change can be used as a way to cool down buildings or streets.\cite{brophy2012green}

Indigenous architecture offers some interesting examples of optimized heat transfers within a house. Windcatchers, windtowers or solar chimneys houses in the Middle-East suck the outside air in, cool it down either by making the draft go through a cooler underground pipe or by putting it in contact with water, before letting it enter the dwelling.\cite{jomehzadeh2017review, abdeen2019solar} Yakhchals in Iran, ice houses in the gardens of the Château de Versailles (France) or in the Boboli Gardens in Florence (Italy) were built to store ice during summer. Yakhchals also had a neighboring pool where ice could be formed at night, thanks to the radiative transfer from the pool to the sky \cite{pochee2017new, Yakhchal} (see the \textit{Temperature within a Yakhchal} problem \cite{SupplMat}).

We finish by discussing the Urban Heat Island effect (see the \textit{Effect of the albedo} activity \cite{SupplMat}) and the effect of vegetation to mitigate it (see the \textit{Cooling effect of trees} activity \cite{SupplMat}).



\section{\label{sec:project}Student projects}
The second part of the course is dedicated to student projects \cite{comment_Projects}. In our curriculum, this is often the first university-level science project students have to undertake. Pedagogically, this part presents all the benefits of project-based learning: an increased student involvement, learning to manage the different parts of a project, learning to anticipate and adapt, learning about the methodology of research and more generally of scientific reasoning, reading literature, and developing manual skills.
The first learning objective is to learn the basics of environmental physics, as described in Section~\ref{sec:cours}.
Our second learning objective for this course therefore is for students to learn the methodology to approach a physics problem: identify a problem, do a literature search, define the means of investigation, and overcome obstacles that may arise in the course of the study. 
Moreover, lectures on climate change often focus on assessing the reality of global warming, and explaining the causes of this tremendous change in our daily environment, without offering practical solutions to mitigate these effects. As a consequence, students may feel that the answers to climate change escape them, thus leading to a lesser involvement in these issues, and to increased distancing and anxiety. We believe these projects offer a partial remedy to this by showing the students they can use their skills in physics to develop a meaningful project in a short time frame (12 hours in class). 

\subsection{Methodology}
For these projects, students choose to work alone or in groups, although we encourage them to team up in groups of up to four. They can investigate any topic in relation to the theme of the course through numerical simulations, case studies, review of the literature, or experiments. 

Although we give examples of possible subjects, we insist  that the definition and decision-making process lie in the students' hands. As choosing and defining a project takes time, we start early on, during the first part of the course, with a collective brainstorming to list possible topics for the projects, although 
some students come to the first lecture with a precise and well-thought-out project that reflects their personal interest.
We then ask them to fill in a Project Description form where they have to list the necessary or required steps towards the completion of the project. They also need to say whether they intend to visit some building or laboratory sites. This encourages them to look beyond the strict boundaries of their curriculum to seek advice and knowledge from academics and/or architects, which they often imagine as out of reach. If their project is experimental, they also need to draft an experimental protocol and list the material, sensors and apparatus they might need.

We deliberately want to keep the projects low-cost and do-it-yourself as much as possible to encourage creativity and have the students realize that they do not need expensive equipment to produce scientific results. For this, the projects benefit from the existing lab infrastructure at the Physics Department of University Paris-Saclay. We have bought some dedicated sensors specifically for this course, including some Arduino-based sensors (humidity level sensors, light sensors, pressure sensors, thermometers,\ldots), that are listed in Section V of the supplementary material.\cite{SupplMat}
Students also make use of consumables such as wood, metal plates, cardboard, or polystyrene.
If a project requires additional materials or equipment, we have a small annual budget of about U.S. \$500 to buy the missing items.

In the next step, the instructors review the various Project Description forms. The objective is to narrow down topics that may be too broad to cover in a 12 hour-project, or to caution the students when we believe a project may be difficult to complete successfully.

The various groups then work autonomously (but under our supervision) on their project, for six weeks. Each week, there is an open session of 2 hours for lab work and discussions with educators. 
For experimental projects, the lab is accessible outside class hours.
For modeling and case study projects, students complete their work at home in addition to these sessions. 
At the end of the course, they have to produce a written report, which they also defend orally. If time permits, we try to organize a separate ungraded presentation session, so that all the class benefits from the projects of the others. Generally, students do quite well on this part: out of the 40 students we have taught, the average grade was 13.6/20 (standard deviation 2.8).

\begin{table*}[t!]
\begin{ruledtabular}
\begin{tabular}{p{0.22\linewidth}p{0.55\linewidth}p{0.05\linewidth}p{0.05\linewidth}p{0.05\linewidth}p{0.05\linewidth}}
&& CS & LR & S & E\\
\hline
\multirow{2}{*}{Climate change} & The influence of glaciers on climate change$^\dagger$  & & \checkmark & & \checkmark\\
& The importance and consequences of measuring temperature &&\checkmark&&\\

\hline
\multirow{3}{*}{Energy production} & Extracting energy from ocean waves$^\dagger$ &&\checkmark&&\\
& Extracting energy from waves (2) &&&&\checkmark\\
&Small wind turbine model&&&&\checkmark \\
& Offshore wind farms &&\checkmark&&\\
& Study of thermoelectric generators for powering mobile devices &&&&\checkmark\\
& Characterization of a thin film solar panel &&&& \checkmark\\ 
& Experimental investigation of the stack effect &&&& \checkmark\\

\hline
\multirow{5}{*}{Energy transfer} & Thermal conductivity of popcorn bricks$^\dagger$ &&&& \checkmark\\
& Thermal conductivity of a cork wall &&&& \checkmark\\
& Thermal conductivity of a straw wall &&&& \checkmark\\
& Heat transfer mechanisms in a house mock-up &&&& \checkmark\\
& Subground housing$^\dagger$ &&&\checkmark&\checkmark\\
& Efficiency of vegetated roofs for heat transfer from the roof$^\dagger$&&&\checkmark&\\
& Insulating materials and application to the case of a Tiny House$^\dagger$ &\checkmark&\checkmark&&\\
& Construction and characterization of a solar oven &&&&\checkmark\\

\hline
Performance assessment & Case study: Parans Light Collector for bringing daylight in dark places$^\dagger$&\checkmark&&&\\

\hline
\multirow{4}{*}{Building performance} & Energy performance of Powerhouse Kj\o rbo (Sandvika, Norway)$^\dagger$&\checkmark&&\checkmark&\\
& Energy and Water consumption performance of ABC Residence (Grenoble, France)&\checkmark&&&\\
& Criteria for a sustainable building &&\checkmark&\checkmark&\\
& Energy efficiency of an apartment - simulations and measurements$^\dagger$&&&\checkmark&\checkmark\\

\hline
\multirow{3}{*}{Urban development} & Simulation of traffic jams$^\dagger$ &&&\checkmark&\\
& Statistical physics modeling of city growth &&& \checkmark&\\
& Modeling of the transportation network of the Paris (France) region &&&\checkmark&\\
& Influence of the sand on African roads$^\dagger$&&\checkmark&&\\
& Efficiency of sand barriers for the control of desertification &&\checkmark&&\\

\hline
\multirow{2}{*}{Water management}  &Water self-sufficiency - consumption and supply methods &&\checkmark&\checkmark&\\
& Water management - Case study of the Plateau de Saclay (France)&\checkmark&\checkmark&&\\

\hline
\multirow{4}{*}{Agriculture and vegetation} & How do plants affect the air we breathe$^\dagger$ &&&&\checkmark\\
& Influence of vegetated schoolyards on the urban heat island effect $^\dagger$&\checkmark&\checkmark&&\\
& Can a region be self-sufficient in vegetables?$^\dagger$  &&\checkmark&&\\
& Ground humidity as a function of ground cover &&&&\checkmark\\
& Light pollution: causes, consequences and solutions &&\checkmark&&\\

\hline
Waste& Using magnetic levitation to sort plastic wastes&&&&\checkmark\\

\end{tabular}
\end{ruledtabular}
\caption{\label{tab:table2}%
List of student projects developed over the course of 3 academic years. CS stands for case studies, LR for literature review, S for simulations, E for experiments. The projects marked with a $^\dagger$ were performed in lock-down during the COVID-19 pandemic.}
\end{table*}

\subsection{Examples of student projects}
The projects developed by students since the creation of this course in spring 2020 are listed in table \ref{tab:table2}. The corresponding reports are accessible on the course's webpage.\cite{projects_archi} As can be seen, they are very diverse and tackle topics related to both climate change and to the built environment. They are relatively balanced between experimental projects, modeling and literature surveys.

\begin{figure}[t]
\centering
\includegraphics[width=0.6\columnwidth]{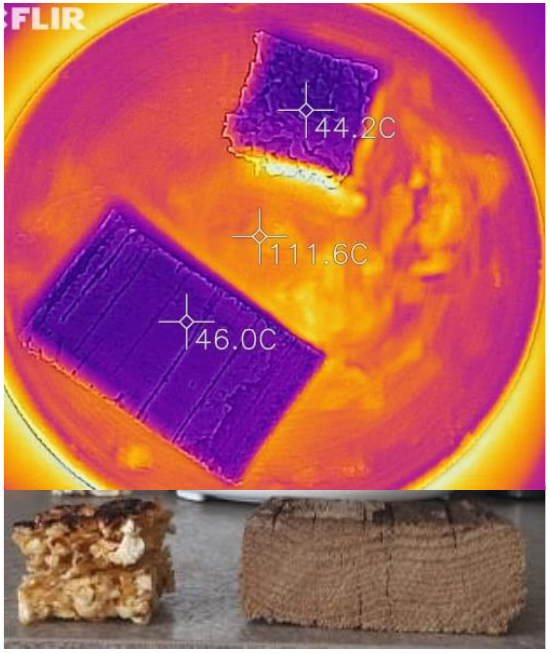}
\caption{Experimental project aimed at comparing the heat conductance of a  popcorn brick to that of a piece of wood. The experiment was performed at home during the COVID-19 lockdown. Courtesy of Th\'eophile Tanguy. The top panel shows an infrared image of the popcorn brick (top black square) and of the piece of wood (left black rectangle) when heated by a frying pan.}
\label{fig:popcorn}
\end{figure}

\begin{figure}[t]
\centering
\includegraphics[width=\columnwidth]{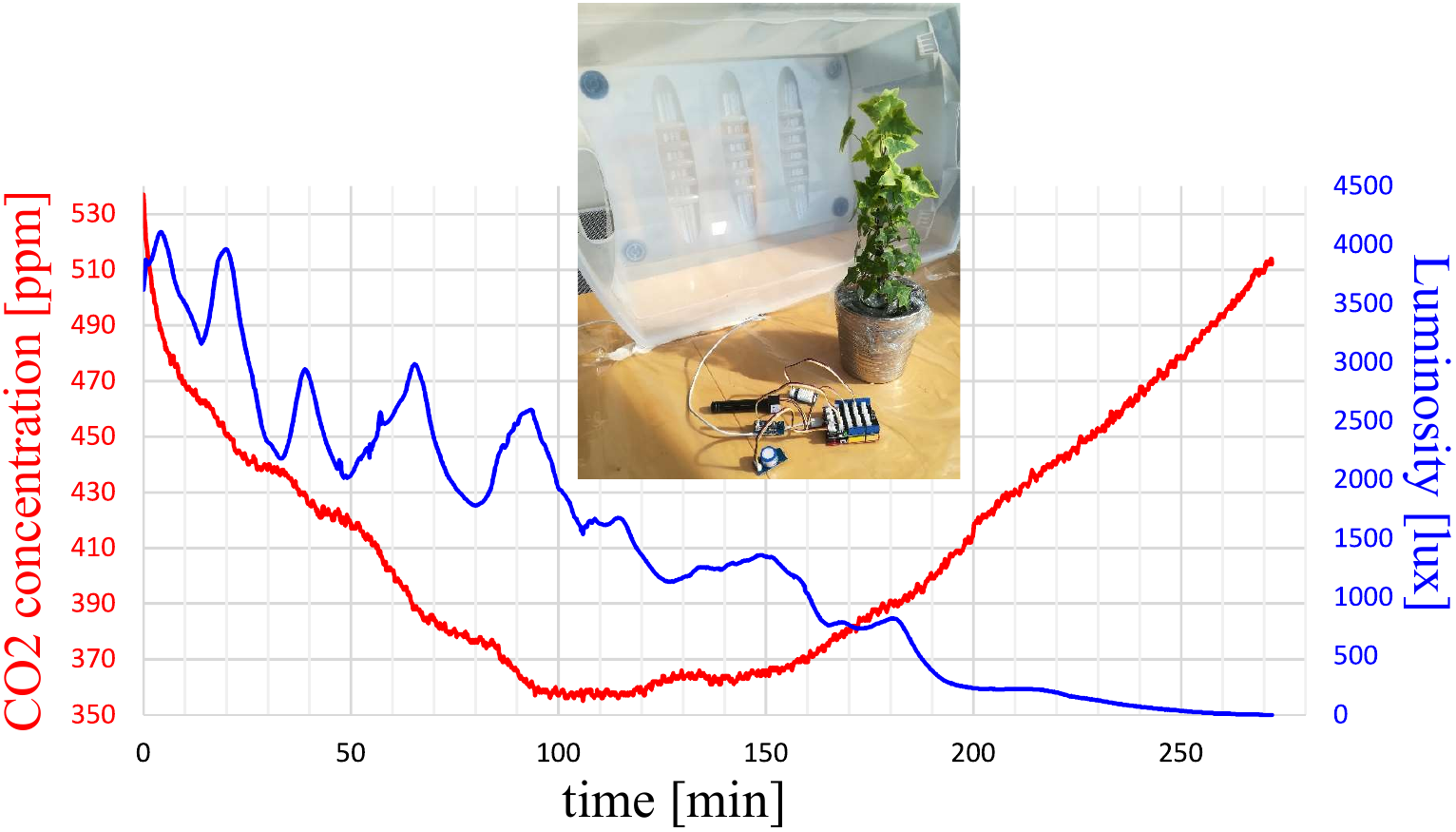}
\caption{Experimental project to correlate the CO$_2$ concentration in a plastic box containing a plant (shown in the inset) with the luminosity. The photosynthesis-dominated phase (when the lighting level is high) gives way at night to a respiration-dominated phase. The experiment was performed at home during the COVID-19 lockdown. Courtesy of Audrey Goutard.}
\label{fig:plant}
\end{figure}

As educators, we have generally found that the students were engaged and enthusiastic when undertaking their project. From the feedback of 31 students, 100\% rated the project part of the course with a grade of 3 or more out of 5. 77\% rated it with the maximum grade of 5. They liked the hands-on experience and the opportunity to investigate a question that interested them. While 55\% of the students who answered the post-course evaluation were satisfied with the balance between lectures and the project, 39\% would have preferred to have more time for their project. 
To us, this shows the importance of having students develop their own reasoning, but also signals an increased level of involvement in projects. Most of those who conducted experimental work were extremely satisfied with the outcome, and this was especially true during the COVID-19 pandemic when they realized they could perform meaningful experiments at home (e.g. assessing the thermal properties of a popcorn brick (figure \ref{fig:popcorn}), measuring the influence of plants' respiration and photosynthesis on the air composition (figure \ref{fig:plant}), and using their own apartment for conducting real-scale experiments on its thermal performance). Students also realized, as it had been emphasized in the Climate Change lectures, the multidisciplinary approach inherent to this topic. 
One group, for instance, collaborated with a geophysicist to analyze the depth-dependent soil temperature that he had measured.\cite{Troglo} Finally, we believe that these projects encouraged students to take action and develop their own local solutions to fight climate change. Several students chose to continue studies of environmental science.
Another has developed a simulation showing how a neighborhood could achieve water self-sufficiency by collecting rainwater. She is currently building a water-retaining tank at her parents' house to achieve this goal.

\section{\label{sec:discussion}Discussion and Conclusion}
The development of this course, with its relatively unusual structure, has been and still is a work-in-progress. We see several benefits, but also a few shortcomings which should be acknowledged.

Starting with the shortcomings, including such dense lectures as well as student projects is probably too ambitious for 12 sessions (overall course time of 24 hours). Students sometimes complain that we cover too many topics and that they lack time for the project.
Another shortcoming is that we do not give an in-depth study of some physics-related topics. 
However, we envision this as a bridge towards specialized graduate level courses. For instance, although we try to convey many non-trivial concepts to students, explaining the equations and techniques at the heart of climate modeling is beyond our expertise and would fill a whole course. 
On the positive side, this intermediate level lecture with a broad spectrum that includes building design is a great opportunity for us to discuss everyday life and the practical aspects of physics, which are often left aside in fundamental physics courses. 
This is also an opportunity to use and mix various tools taken from different physics courses as well as from other fields.
We believe this makes the course formative for students well beyond environmental physics.
Last, such a course has also been very formative for us, since it challenged our understanding of physics through a wide range of applications.
Preparing and supervising this course is a great source of surprises, questions and motivation. 

\section*{Author Declarations}
The authors have no conflicts to disclose.

\acknowledgments
G.R. thanks C. Even and G. Blanc for many fruitful discussions.
We also thank all students from the Physics of Sustainable Development course at Paris-Saclay University for their work, their ideas, their enthusiasm, and their feedback. Finally, we thank P. Puzo for giving us the opportunity to develop this course within the Magist\`ere de Physique Fondamentale curriculum.


\bibliography{apssamp}

\begin{thebibliography}{72}
\bibitem [{cli()}]{climate_activity}%
  \BibitemOpen
  \href@noop {} {}\bibinfo {howpublished} {Each group then presents its
  findings to the class. The aim of this activity is to have them compare
  different climates, while learning to look for meteorological data and to
  present them in a clear and meaningful way.}\BibitemShut {Stop}%
\bibitem [{com({\natexlab{a}})}]{comment_climate_assessment_activ}%
  \BibitemOpen
  \href@noop {} {}\bibinfo {howpublished} {During this activity, students often
  choose to investigate the hometown of one group member. When they present
  their work to the class, this results in a large variety of locations, which
  is a way of promoting diversity.} ({\natexlab{a}})\BibitemShut {NoStop}%
\bibitem [{Ene()}]{Energy_sources}%
  \BibitemOpen
  \href@noop {} {}\bibinfo {howpublished} {We restrict outselves to locally
  implementable energy sources, and chose not to discuss hydroelectric or
  nuclear power for instance.}\BibitemShut {Stop}%
\bibitem [{com({\natexlab{b}})}]{comment_Projects}%
  \BibitemOpen
  \href@noop {} {}\bibinfo {howpublished} {During the projects, students often
  like to tackle a subject that is relevant in their life. They are sometimes
  inspired by building techniques or sustainable solutions that are used in
  their home countries or regions. Explaining this to their classmates favors
  intercultural exchanges.} ({\natexlab{b}})\BibitemShut {NoStop}%
\end{thebibliography}

\newpage
\appendix

\onecolumngrid

\begin{center}
\noindent{\Large \textbf{A course on Climate change and Sustainable Building Design - Supplementary Materials}}
\end{center}

\section{Examples of class activities on common concepts}

\subsection{The Solar constant}
The solar constant is the power per unit surface emitted by the Sun at the Earth's distance.

\noindent \textbf{Question: Assuming the Sun is a perfect blackbody, determine the power that is emitted by its surface.}

\noindent \textbf{Answer}: The total power emitted by the Sun is $P_{\rm total}=\oiint \sigma T_{\rm Sun}^4 dS=4\pi R_{\odot}\sigma T_{\rm Sun}^4\simeq3.9\times 10^{26} W$ with $\sigma$ the Stefan-Boltzmann constant, $R_{\odot}$ the Sun's radius and $T_{\rm Sun}$ its temperature.

\noindent \textbf{Question: Assuming that space in between the Sun and the Earth is only filled with vacuum, calculate the solar constant.}

\noindent \textbf{Answer}: We assume the Sun emits radiation isotropically. Then, the power per unit surface received at a distance of $D=1.5\times 10^8$ km is $P_{\rm Earth}=P_{\rm total}/(4\pi D)=1376$ W.m$^{-2}$. This value is called the solar constant which can vary annually and also depends on the solar cycle. This roughly represents the power we receive from the Sun during daytime. The earthen surface intercepting the Sun's radiation is $S=\pi R_{\rm Earth}^2\simeq1.27\times 10^{14}$ m$^2$, with $R_{\rm Earth}$ the Earth's radius, so that the total power received by the Earth can be estimated to be $P_{\rm received}\simeq 1.7\times 10^{17}$ W.

\noindent \textbf{Question: Compare this value to the energy consumption of a typical French household (4.77 MWh/yr).}

\noindent \textbf{Answer}: If 1 m$^2$ or earthen surface receives 1376 W, the energy consumption of a typical French household represents the energy received by 10 m$^2$ for 15 days (assuming 24 hours illumination). Alternatively, it represents the energy received by 0.4 m$^2$ (constantly illuminated) for one year. (One could play around with the figures, for instance by taking into account the hours of daylight in a given location).

\subsection{Climate assessment}
\begin{enumerate}
    \item Work in groups of 2 or 3.
    \item Choose a location.
    \item Establish its climate using the Methodology described below.
    \item If you were to build a house or a building on this location, what are the parameters you would keep in mind during the project design?
\end{enumerate}

\paragraph*{Methodology to determine the site's climate:}
\begin{itemize}
    \item Download the Excel file used to visualize the meteorological situation of a site (\url{https://www.dropbox.com/s/xs1iq5rzmy33v0h/Meteorological_Conditions_simplified.xlsx?dl=0})
    \item Download the meteorological data of your chosen site by going to the Iowa Environmental Mesonet (\url{http://mesonet.agron.iastate.edu/request/download.phtml}). 
    \begin{itemize}
        \item Select the Network corresponding to the country the city belongs to and the Station closest to the city.
        \item Select ``All Available'' in the step 2.
        \item In step 3, select the desired date range. Warning: beyond 2 years data range, the calculations may take a long time.
        \item Select ``Coordinated Universal Time'' in step 4.
        \item In step 5, use the following parameters:
        \begin{itemize}
            \item Data format: ``Tab delimited (no DEBUG headers)
            \item Include Latitude + Longitude: ``No''
            \item How to represent missing data: ``Use blank/empty string''
            \item How to represent Trace reports: ``Use blank/empty string''
            \item ``Save result data to file on computer''
            \item Leave the default option in step 6.
        \end{itemize}
        \item Click on ``Get Data'' in step 7.
    \end{itemize}
    \item Open the Excel file and go to the ``Data'' worksheet. Suppress all data if this worksheet is not empty and replace it with the data from the downloaded file. Check that the decimal separator is ``.'' and not ``,''.
    \item Go to the ``Precipitation\_data'' worksheet and look for the weather station closest to your site. Determine the line number corresponding to the nearest station. If several occurrences of the same weather station appear, look at the corresponding values in column G ``Statistic Description'' and choose the line corresponding to ``Mean Monthly''.
    \item Click on ``Calculate''. Warning: The calculation may take a few minutes. You have now set everything you need for the different graphs that are in the various worksheets.
\end{itemize}       

\paragraph*{Methodology to determine the site's solar irradiation:}
\begin{itemize}
    \item Go to the Global Solar Atlas (\url{https://globalsolaratlas.info/map}) and select your site.
    \item Scroll down the right panel and click on ``Open detail''.
    \item You can download the information in an Excel file.
\end{itemize}

\subsection{Cooling effect of trees}

\noindent \textbf{Question: One large tree can produce 450 L of water per day. Knowing that the latent heat of evaporation is $L_{\rm evap}=2'265$ kJ/kg, estimate the cooling power of a tree and compare it to the one of a domestic cooler ($\sim$ 2.5 kW).}

\noindent \textbf{Answer}: The energy corresponding to the evaporation of 450 L is $E=1'019$ MJ. This energy is the daily available cooling energy. It corresponds to a power of $P=11.8$ kW. This energy is equivalent to 6 domestic coolers running for 19 hours each day. This highlights the enormous cooling power of the vegetation. 


\section{Examples of class activities for the lectures on climate change}

We give below a few examples of class activities. Students usually search these at home or during the classroom with the help of internet. This explains why the questions are rather open and data usually not given.

\subsection{The carbon conversion factor}
\noindent \textbf{Question: Estimate how many gigatons of \CO2 one has to emit to increase by 1 ppm its concentration in the atmosphere (not taking into account that it is partly reabsorbed in the ocean and the biosphere).}

\noindent \textbf{Answer}:
In order to find the order of magnitude, we need to find the mass of \CO2 corresponding to a concentration of $c=$ 1ppm in the atmosphere. We have $c=10^{-6}=n_{\CO2}/n_{\text{air}}$ where $n_i$ are number of moles. 
Masses and moles are related by molar masses $M_i$. From $m_{\CO2} = n_{\CO2} M_{\CO2}$ and $m_{\text{air}} = n_{\text{air}} M_{\text{air}}$ we get
\begin{equation}
 m_{\CO2} = 10^{-6}\frac{M_{\CO2}}{M_{\text{air}}} m_{\text{air}}.    
\end{equation}
From chemistry, we know that $M_{\CO2} = 12+2\times 16 = 44$ g.mol$^{-1}$ and 
$M_{\text{air}} \simeq 78\%\times M_{\N2}+21\%\times M_{\O2} \simeq 29$ g.mol$^{-1}$. Now, to estimate the mass $ m_{\text{air}}$ of the atmosphere, we use that pressure $P$ over the Earth surface at the sea level approximately corresponds to the weight of air per square meters, ie. $P = 10^{5}$ Pa $ = m_{\text{air}}g/ S$ with $g\simeq 9.8$ m.s$^{-2}$ and $S=4\pi R^2$ with $R=6370$ km. One finds $m_{\text{air}} \simeq 5.2\,10^{18}$ kg with this estimate. Finally, one finds $m_{\CO2} \simeq 7.8$ Gt (gigatons).
The actual number given in the lecture is 7.3 Gt of \CO2 increases by 1 ppm the atmospheric concentration of carbon dioxide. Accurate and up-to-date figures about the carbon emissions and carbon cycle can be found in
\url{https://www.globalcarbonproject.org/}.

\subsection{Radiative balance of the Earth and retroaction effects}

This activity is done after an introduction to the radiative balance of Earth.
Notations and reasoning are strongly inspired by the book \cite{Delmas2007}.\\

\textbf{Question: One shell atmospheric model. Find the equilibrium temperature of the Earth surface temperature $T_S$ submitted to average incoming solar flux $\mathcal{F}_s\simeq 342$W/m$^2$ and in the presence of atmospheric albedo $\mathcal{A}_b$, absorption of visible light $\alpha$, absorption and re-emission of infra-red light with coefficient $\epsilon$ by atmosphere at temperature $T_a$.}

\textbf{Answer:} Writing radiative balance at three levels
\begin{align*}
\text{sky: }&\mathcal{F}_s = \mathcal{F}_s\mathcal{A}_b + (1-\epsilon)\sigma T_S^4 + \epsilon\sigma T_a^4\,, \\
\text{atmosphere: }&\mathcal{F}_s + \sigma T_S^4  = \mathcal{F}_s(1-\mathcal{A}_b)(1-\alpha) + \mathcal{F}_s\mathcal{A}_b  +(1-\epsilon)\sigma T_S^4 + 2\epsilon\sigma T_a^4 \,, \\
\text{surface: }&\mathcal{F}_s(1-\mathcal{A}_b)(1-\alpha) + \epsilon\sigma T_a^4 = \sigma T_S^4\,.
\end{align*}
with $\sigma$ the Stefan-Boltzmann constant.
Simplifying these equations gives
\begin{equation}
    T_S = \left(\frac{1-\alpha/2}{1-\epsilon/2}\right)^{1/4} T_{\text{eq}}
    \;\text{with}\; T_{\text{eq}} = \left(\frac{\mathcal{F}_s(1-\mathcal{A}_b)}{\sigma}\right)^{1/4}.
\end{equation}

\textbf{Question: We write $\phi$ the outgoing radiative flux to the sky. The net flux at the top of troposphere is $N=\mathcal{F}_s - \phi$ with $N=0$ at equilibrium. 
This net flux $N$ depends on parameters that we consider external to the climate system and denoted by $\vec{E}$ and internal $\vec{I}$, including the surface temperature $T_S$ that we single out. The internal parameters are assumed to be function of $T_S$. After writing an elementary variation $\Delta N$ w.r.t its parameters, show that the surface temperature increase after reaching equilibrium takes the form
\begin{equation}
    \Delta T_S^{\text{eq}} = \frac{G_0}{1-RG_0} \Delta F.
\end{equation}
}

\textbf{Answer:} We start from
\begin{equation}
\Delta N = \sum_i \pdv{N}{E_i} \Delta E_i + \left(
\pdv{N}{T_s } +\sum_j \pdv{N}{I_j}\pdv{I_j} {T_S}
\right)\Delta T_S.
\end{equation}
After the variation, the system gets back to equilibrium where $\Delta N = 0$. We get $ \Delta T_S^{\text{eq}} = G \Delta F$, with $\Delta F = \sum_i \pdv{N}{E_i} \Delta E_i$, the sum of radiative forcings, in W/m$^2$. Example of external radiative forcing can be a change of incoming solar flux $\mathcal{F}_s$ or an anthropogenic source of greenhouse gases.
The total gain decomposes into
\begin{equation}
    G^{-1} = \underbrace{-\pdv{N}{T_S}}_{G_0^{-1}} - \underbrace{\sum_j  \pdv{N}{I_j}\pdv{I_j} {T_S}}_{R}.
\end{equation}
$G_0$ is the bare gain, or direct response of the surface temperature w.r.t. radiative forcing. $R$ is called retroaction and quantifies the indirect response. If $T_S$ changes, $T_j$ changes so as to increase or decrease the response w.r.t the bare gain. Within our convention, $R>0$ means positive retroaction with $G>G_0$ while $R<0$ means negative retroaction $G<G_0$. This essentially equivalent to the amplifier principle and an analogy with electronic can be made.\\

\textbf{Question: Taking the simplest model without greenhouse effect where $N =\mathcal{F}_s(1-\mathcal{A}_b) -\sigma T_S^4$. Compute $G_0$ and $R$ assuming that $\mathcal{A}_b$ is a function of $T_S$ only. Discuss the expected sign of the retroaction for sea ice melting.}

\textbf{Answer:} Applying the definitions, we find
\begin{equation}
 G_0 = 1/(4\sigma T_S^3)\; \text{ and }\; R = - \mathcal{F}_s\pdv{\mathcal{A}_b}{T_S}.
\end{equation}
we see that $G_0 > 0$. $\mathcal{A}_b$ is an example of $I_j$ variables. For $R$ and sea ice response, we can say that albedo $\mathcal{A}_b$ decreases when $T_S$ increases because sea ice reflects visible light much more than water. Therefore,  $\pdv{\mathcal{A}_b}{T_S}<0$ and $R>0$. This is a positive retroaction that enhances even further the increase of surface temperature, which is quite intuitive.\\

\textbf{Question: We consider the out-of-equilibrium response in a very naive model.
We consider that the climate system has a global thermal capacity $C$ per $m^2$ such that the internal energy is $C\Delta T_S$ up to a constant. Discuss the dependence of the transient time with the climate system parameters.}

\textbf{Answer:} The variation of internal energy equates the variation in net flux so we have
\begin{equation}
C\dv{\Delta T_S}{t} = \Delta N = \Delta F - G^{-1} \Delta T_S \;.
\end{equation}
The solution is $\Delta T_S(t) = \Delta T_S^{\text{eq}} (1-e^{-t/\tau})$ in which $\Delta T_S^{\text{eq}}$ is the equilibrium temperature shift discussed above and $\tau = GC$ is the transient time. We thus see that the gain and the thermal inertia govern the temperature increase in the transient regime. 
These qualitative discussions can be completed with the reading of Ref.~\cite{hansen1981climate}, which is readable for students and plunge them into the history of climate modeling.

\subsection{Power and \CO2 emissions of thermal car}

 \textbf{Question: Try different ways of estimating the average / typical power of a car engine.}

 \textbf{Answer}:
We need to divide a typical energy by some time by considering a typical transformation. The simpler one is when the car starts from 0 velocity to 100 km/h with full power of the engine.
If you take a regular small car of mass $M\simeq 1000$ kg, that reaches $v=100$ km/h in about $t=10$ seconds (figures can be found on the internet) it yields $P = \frac{1}{2}Mv^2/t = 0.5\times10^3\times(100.10^3/3600)^2/10 \simeq 40$ kW.
Notice that thermal car engine are usually given in horsepower unit with the typical order of magnitude  100 hp $\simeq$ 74 kW.

Another strategy, better adapted to average power, is to look to the energy content of gasoline, typically $1L \to 9$kWh (see eg. \url{https://en.wikipedia.org/wiki/Gasoline}).
Then, if you take the example of driving in cities at $25$ km/h on average and consuming 8L/100km, it means you burn 2L/hour corresponding to an average power 
$P \simeq 2\times 9 \simeq 18$ kW. Interestingly, electrical cars now directly show these order of magnitudes in kW on the counter.\\

\textbf{Question: Estimate the number of kg\CO2 that gasoline emits per liter. For this, consider burning octane $\ce{C8H18}$, given that 1 L of gasoline weights 750 g.} 

 \textbf{Answer}:
The composition of gasoline is complex but essentially contains hydrocarbons that are alkanes, among which octane $\ce{C8H18}$ can serve as a reference for the estimate. Its molar mass is $M_{\ce{C8H18}} = 8\times 12 + 18 = 114$ g/mol.
Thus in 1L of octane, one has $n = 750/114 = 6.58$ mol.
Octane is burned with oxygen through the chemical reaction with overall balance
\begin{equation}
{\ce {C8H18 + 12.5 O2 -> 8 CO2 + 9 H2O}}    
\end{equation}
in which one mole of octane gives 8 moles of \CO2. Consequently, using $M_{\CO2} = 44$ g/mol, the emitted mass of \CO2 that is emitted from burning 1 L of gasoline reads
\begin{equation}
m_{\CO2} = 8\times n \times 44 \simeq 2.3\, \text{kg}
\end{equation}
More precise figures can be found on the internet but they are all close to this rough estimate.\\

\textbf{Question: Infer the typical range of emission in g/km as given on car labels when you buy a car. 
Estimate a/your typical car emissions over a year and compare with worldwide \CO2 emissions. Estimate the emissions per energy in g\CO2/kWh.}

\textbf{Answer}:
If one considers 7 L/100 km as a typical car consumption, it makes $7\times 2.3\,10^3 / 100 \simeq 150$ g/km, which is what one can find the car labels.
Taking again 9 kWh/L, the emissions per kWh are around 210 g\CO2/kWh. 
If you fill up a tank of 40 L each 2 weeks, this makes roughly 1000 L/year and 2.3 t\CO2.
We recall that the global emission are about 36 Gt\CO2/yr, thus corresponding to about $36\,10^{9}/(7.5\,10^{9}) \simeq 5$ t\CO2 per person per year, equivalent to burning 2000 L of gasoline.

\subsection{Demography modeling and societies}

The motivation for this activity, which was actually never given in this depth by lack of time (although the main conclusions about demography are given), is that it is an important subject, supported by extensive data, and benefitting from nice online simulators. Therefore, population modeling is often a social science topic with which physicists feel comfortable with, scientifically speaking. One important and interesting thing is that the deeper the modeling, the more social aspects are revealed. The goal is to guide students towards demographers' conclusions about population growth, which is essentially that it is reaching a plateau, that aging will be the leading challenge in western countries and that the global population is not a relevant parameter for environmental challenges, although the ``extensivity'' of impacts might at first look like an attractive argument for a natural science-trained person.

\subsubsection{Naive views}

\textbf{Question: The total population growth rate $\alpha = \frac{1}{N(t)}\dv{N}{t}(t)$ is constant in the Malthus (1826) model, while it follows $\alpha = \alpha_0(1 - N/N_{\infty})$ with  $\alpha_0 > 0$ and $N < N_{\infty}$ in Verhulst (1840) model.
Let $N_0$ be the initial population, find the time evolution of each model and discuss its behavior.}

\textbf{Answer:} Malthus gives an exponential increase $N(t) = N_0e^{\alpha t}$ assuming a positive growth rate. Although, such an exponential growth can qualitatively reproduce some recent population data over a finite range of time. It is clearly unrealistic on the long run. Verhulst's model possesses an intrinsic regulation term with $N_{\infty}$ a maximum population. Its solution is a logistic function
\begin{equation}
N(t) = \frac{N_\infty}{1 + \frac{N_\infty-N_0}{N_0}e^{-\alpha_0 t}}.
\end{equation}
These are classic examples of simple dynamical systems with trivial fixed points. The logistic equation is also important for finite resource peak modeling. 
Yet, these models are totally uninformative about the true parameters governing a realistic population dynamics.

\subsubsection{Modeling population pyramid}

\textbf{Question: Let $N_a(t)$ be the number of people of age $a=0,1,2,\ldots$ in year $t$, so that $N(t) = \sum_{a=0}^{120} N_a(t)$ is the total population.
The distribution $N_a(t)$, formally put in vector form $\vec{N}(t)$, is the population pyramid usually split between men and women. Let $\tau_a(t)$ the rate of birth in the population slice $a$ in year $t$.
Let $s_a(t) = \frac{N_{a+1}(t+1)}{N_a(t)}$ the survival probability of population of age $a$, so that $d_a(t) = 1-s_a(t)$ is the corresponding death rate.
Discuss the typical form for the functions $d_a$ and $\tau_a$ as a function of $a$. Is there a typical time scale ?
Write in a matrix form the evolution equation $\vec{N}(t+1) = \mathbf{A}(t) \vec{N}(t)$. \\
Using Ined online simulators, observe the expected UN evolution and play with parameters to see the effects over population growth:\\
\url{https://www.ined.fr/en/everything_about_population/population-games/tomorrow-population/}}

\textbf{Answer: } The form of birth rate $\tau_a$ is expected to be a dome starting around $a=15$ towards $a=45$ for biological reasons. Then, the actual shape depends on societies and its features, themselves evolving with time $t$.
Clearly, this sets the ``generation'' time scale, varying between 20 to 30 years depending on societies.
The form of death rate is usually a dip (for infancy deaths) at short age towards $a=15-20$ followed by an increase towards almost 1 at large ages. Here also, the form and the scales are strongly dependent on societies.
Examples of curves can be fund in demography textbooks, such as Ref~\cite{lundquist2015demography} or online national statistics websites.
The evolution equation is simply
\begin{equation}
\begin{pmatrix}
N_0(t+1)\\ N_1(t+1)\\ \vdots \\ \vdots \\ \vdots\\ N_{120}(t+1)
\end{pmatrix}
= \begin{pmatrix}
\tau_0(t) & \tau_1(t) & \tau_2(t) & \tau_2(t) & \cdots & \tau_{120}(t) \\
s_0(t) & 0 & 0 &0 &  & 0\\
0 & s_1(t) & 0 &0 &  & 0\\
0 & 0 & s_2(t) &0 & \ddots & 0\\
\vdots & \vdots & \ddots &\ddots & \ddots & \vdots\\
0 & 0 & 0 &0 & s_{121}(t) & 0
\end{pmatrix}
\begin{pmatrix}
N_0(t)\\ N_1(t)\\ \vdots \\ \vdots \\ \vdots \\ N_{120}(t)
\end{pmatrix}.
\end{equation}
This structure and essential definitions help students better understand essential demographic parameters. It also shows that such evolution is tracked by life tables gathered by demographers.
This is the structure and parameters that are at the heart of population projections.
For instance, the probability to die after $y$ years when one has age $a$ reads 
\begin{equation}
    P_a(y) = s_a\times s_{a+1}\times\cdots\times s_{a+y-1}(1-s_{a+y}),
\end{equation}
which roots the not-so-intuitive definition of life expectancy at age $a$
\begin{equation}
LE(a) =\sum_{y=1}^{120-a} y P_a(y).
\end{equation}
with $LE(0)$ the life expectancy at birth, most widely discussed. These concepts are essential to understand how projections are built and one can play with online simulations to understand the typical evolution.
These simulations illustrate well concepts such as demographic inertia, demographic transition, demographic crash, or population aging.

After this formal background, the discussion can naturally switch towards what controls these functions and parameters. For this, one must refer to demographers textbooks, see eg. Refs.~\cite{harper2018demography, veron2013demographie} and show the complexity and systemic social issues that are involved. Eventually, connection with climate change and urbanization challenges can be made to move back to the main topics of the course.


\section{\label{sec:Class_activ_archi}Examples of class activities for the lectures on sustainable building design}
 
\subsection{\label{subsec:climate}Principles of bioclimatic construction}

\subsubsection{Assessing the sustainability labels}
\begin{enumerate}
    \item Choose one sustainability label.
    \item Determine the criterion upon which the certification is based.
    \item Do you think the label adequately reflects what a sustainable building should be?
\end{enumerate}

\subsubsection{Orders of magnitude}
Get in groups of 2 or 3. Choose one of the following physical quantities: power, solar irradiation (direct, diffuse, global), lighting, water consumption, energy, particulate Matter concentration, caloric supply from food, relative humidity, wind velocity, or rainfall. Give a few orders of magnitude. If you make some assumptions, clearly state them. If you base your figures on calculations, give your sources and/or references. If possible, use practical examples. If you can, illustrate them. Summarize your findings in a presentation.

\noindent \textbf{Example:} 1 MWh of electrical energy =
\begin{itemize}
    \item 0.86 tep
    \item 0.000114\% of the annual energy production of a typical French nuclear plant. Source: Wikipedia, average energy produced in 1 year = 9 TWh).
    \item 23.6 m$^2$ of office in France (energy for). Source: Observatoire Immobilier Durable, in France 422 kWh are annually used for heating/cooling/supplying energy/... for 1m$^2$ of office.
    \item 0.00033\% of the energy consumption of French data centers. Source: Sciences et Avenir 2015. All data centers in France use 3 TWh per year.
    \item 0.01 public pools heated. Source: Econergie, an average-sized pool of 500 m$^2$ uses $\sim$ 1000 MWh of energy per year.
\end{itemize}

\noindent The educator can put more or less emphasis on the graphical representation of the data. In our opinion, this is a skill that is very seldom taught in physics class, but is paramount when it comes to communicating the results of your work with others. This applies to researchers presenting their work in conferences, seminars, or grant applications, but also to researchers popularizing their research, or to engineers trying to explain the technical aspects of a project to non-scientists. 

\subsection{\label{subsec:arhi}Energy in buildings}

\subsubsection{Estimation of a Canadian well's efficiency}

\noindent\textbf{Question: Let us consider a typical Canadian well, which brings fresh air into a house through a horizontal PVC pipe. Determine the temperature of the air $T(z)$ as it is sucked along the pipe ($z$ is the pipe axis), considering the following data:}
\begin{itemize}
    \item The surface of the house is 100 m$^2$.
    \item The height of the house is 3 m.
    \item The recommended air renewal rate is of $4V$/h, where $V$ is the volume of the house.
    \item The incoming air is at the external temperature $T_{ext}=35\rm ^o$C.
    \item The Canadian well is surrounded by the soil at temperature $T_{soil}=15\rm ^o$C.
    \item The PVC pipe is $D=$20 cm in diameter.
    \item The thickness of the PVC is $e_{PVC}=$2 mm.
    \item The thermal conductivity of PVC is $\lambda_{PVC}=0.19$ W/m.K.
    \item The density of air is $\rho_{air}=1.2$ kg/m$^3$.
    \item The specific heat of air is $c_{air}=29$ kJ/kg.K.
\end{itemize}
\noindent \textbf{What is the characteristic length on which the incoming air is thermalized to the ground's temperature? If the pipe is 35 m long, what is the temperature of the air when it arrives in the house?}

\textbf{Answer:} We will assume that the air is cooled down via thermal conduction from the pipe to the soil. We will neglect turbulence in the pipe, air conductivity, and all other heat transfer mechanisms. The air renewal rate is of $Q_{air}=4V$/h=1'200 m$^3$/h=0.33 m$^3$/s. Considering the heat balance between $z$ and $z+dz$, one gets:
\begin{equation}
    c_{air}\rho_{air}Q_{air}\left(T(z+dz)-T(z)\right)=\frac{\lambda_{PVC}\pi D dz}{e}\left(T_{soil}-T(z)\right),
\end{equation}
\noindent which gives the following differential equation to solve:
\begin{equation}
    \frac{\partial T}{\partial z}=\alpha \left(T_{soil}-T(z)\right),
\end{equation}
\noindent with 
\begin{equation}
    \alpha=\frac{\lambda_{PVC}\pi D}{e_{PVC}c_{air}\rho_{air}Q_{air}}.
\end{equation}
\noindent which yields the solution:
\begin{equation}
    T(z)=T_{soil}+\left(T_{ext}-T_{soil}\right)\text{e}^{-\alpha z}.
\end{equation}
\noindent The characteristic length $1/\alpha=192$. For a 35 m-long PVC pipe, which is relatively realistic, the air will arrive in the house with a temperature of 31.7$\rm ^o$C. 

\subsubsection{Determining a site's photovoltaic potential}

\begin{itemize}
    \item Go to Global Solar Atlas  (\url{https://globalsolaratlas.info/map}) to determine the optimum orientation of photovoltaic (PV) modules for the site ("Optimum Tilt of PV modules").
    \item Go to Onyx Photovoltaic Estimation Tool (\url{https://www.onyxsolar.com/photovoltaic-estimation-tool}) and estimate the production of PV modules on your site, with the optimum angle.
\end{itemize}

\subsubsection{Wind Power}
\noindent \textbf{Question: Look for the nominal power of a domestic wind turbine and of a large offshore wind turbine. Compare this energy production capability to the energy consumption of a typical French household (4.77 MWh/yr).}

\noindent \textbf{Answer:} There are various kinds of domestic wind turbines, but we will consider a relatively small one which has a nominal power of 1 kW, with a capacity factor (the ratio between the actual output and maximum possible output) of 25\% \cite{WindEurope}. It produces $1\times365\times24\times 0.25=2'190$ kWh each year, which represents about 46\% of a typical French household energy consumption. Note that this reasoning considers the total energy, not the power that can be instantaneously delivered to the household appliances. A large offshore wind turbine typically has a nominal power of 3 MW and has a capacity factor of more than 40\%. It therefore produces about 10 GWh each year, which represents the energy consumption of about 2'200 typical French households.

\subsection{\label{subsec:arhi}Thermal considerations in buildings and
in the city}

\subsubsection{Temperature in an igloo}

\noindent \textbf{Question: Determine the temperature in an igloo. Introduce your assumptions and determine what is the most efficient way to heat the igloo.}

\noindent Data that may be useful: 
\begin{itemize}
    \item Thermal conductivities: $\lambda_{\rm snow}=0.15$ W/m.K, $\lambda_{\rm packed \;snow}=0.3$ W/m.K, $\lambda_{\rm ice}=1.7$ W/m.K
    \item The Stefan-Boltzmann constant: $\sigma=5.67\times 10^{-8}$ W.K$^{-4}$.m$^{-2}$.
\end{itemize} 

\noindent \textbf{Answer:} We will consider the thermal conduction through the igloo's walls. we will examine whether a candle or a person is more efficient in heating the igloo. The power due to thermal conduction between the outside at temperature $T_{ext}$ and the inside at temperature $T_{in}$ can be estimated for a hemispheric igloo of radius $R=2$ m which walls are made of packed snow (estimated $d=$0.5 m in thickness):
\begin{equation}
    P_{\rm cond}=-2\pi R^2 \frac{\lambda_{\rm packed\; snow}}{d}\left(T_{ext}-T_{int}\right).
\end{equation}
\noindent This gives $P_{\rm cond}=-15.07\left(T_{ext}-T_{int}\right)$. This power is counterbalanced by the heating power of either a candle or a single human being. A candle has an emissivity of about $\varepsilon\sim0.9$, a temperature of $T_{\rm candle}=1'000$ K, and a surface of about $10^{-4}$ m$^2$, which emits a power of $P_{\rm candle}=5.1$ W, leading to a temperature difference between the outside and the inside of $\Delta T=T_{int}-T_{ext}=0.3$ K. A person has an emissivity of about $\varepsilon\sim0.5$, a temperature of $T_{\rm person}=310$ K, and a surface of about 2 m$^2$, which emits a power of $P_{\rm person}=524$ W, leading to a temperature difference between the outside and the inside of $\Delta T=35$ K. This can be compared to the Wikipedia entry for igloos \cite{igloo1} which writes that the inner temperature varies between -7$^{\rm o}$C and 16$^{\rm o}$C when the outside temperature is of -45$^{\rm o}$C. The Ice hotel \cite{igloo2} says their rooms have an average temperature of -5$^{\rm o}$C when the outside temperature is -25$^{\rm o}$C. The order of magnitude obtained through this very rough calculation is therefore not so bad!

\subsubsection{Effect of the albedo}
In 2011, New York City (USA) launched a pilot program called the New York ``Cool Roofs'' initiative to mitigate the city's urban heat island effect. Through subsidies, the city encouraged people to paint their roof white (contrary to the traditional dark rooftop materials). 

\noindent \textbf{Question: Assuming a typical composition of a rooftop (for instance, from outside to inside: 7 cm of gravel, an anti-punching separation layer of negligible thickness, 7 cm of thermal insulation, a waterproofing membrane of negligible thickness, and 40 cm of concrete), determine the temperature difference at the roof surface for black gravel or white gravel. Compare your results to the real-life measurements presented in reference \cite{gaffin2012bright}: due to lower albedo, white paint absorbs less energy than black one, and the white rooftops can be 20 to 30$^\circ$C cooler than black ones. }

\noindent Data that may be useful: 
\begin{itemize}
    \item Specific heat: $c_p=4200$ J/kg.K for water, $c_p=1000$ J/kg.K for stone.
    \item Thermal conductivity: $\lambda_{\rm water}=0.6$ W/m.K, $\lambda_{\rm stone}=1.8$ W/m.K.
    \item Density: $d=1000$ for water, $d=2300$ for stone.
\end{itemize} 

\noindent \textbf{Answer:} We will consider that black gravel has an emissivity of $\varepsilon_{\rm black}=0.2$, while the white one has an emissivity of $\varepsilon_{\rm white}=0.9$. New York City has a global horizontal irradiation of the order of $E_{in}=5$ kWh/m$^2$. This means that the black gravel absorbs 4 kWh/m$^2$, whereas white gravel absorbs 0.5 kWh/m$^2$. We will make the crude approximation that this energy is entirely dedicated to heating the gravel: $E_{\rm absorbed}=c\rho d S\Delta T$ where $c$ is the specific heat of stone, $\rho$ the volume mass of stone, $d=7$ cm the thickness of gravel, $S=1$ m$^2$ the considered roof surface, and $\Delta T$ the temperature difference between the outside air and the bottom of the gravel layer. For black gravel, we find $\Delta T = 89$ K, whereas it is $\Delta T=11$ K for white gravel. The graph shown in \cite{gaffin2012bright} shows that the surface temperature of a black surface is 30 K higher than that of a white surface. Once again, although we made very crude assumptions, neglecting the thermal conductivity through gravel or through the building's roof (via the concrete roof), we find the correct order of magnitude to describe the effect of the roof albedo. Of course, finer models could be studied that could also include the wind's cooling effect for instance. The message we would like to convey to students is that a back-of-the-enveloppe calculation can already give you an idea of the magnitude of a phenomenon. In class, we discuss what other phenomena we need to take into account for more precision. Also note that we give more numerical data and information than needed to solve the problem, so that students are (less) influenced by these to consider which phenomena to consider and which ones to neglect.

\section{\label{sec:exams}Other problems taken from exams}

We give below a list of problems given at the exam and evaluating the lectures part of the course. These problems are oversimplification of realistic situation that are addressed with basic, though physical reasoning. We do not claim that the numbers proposed in solutions are reliable but they should be correct orders of magnitude.

\subsection{\CO2 emissions from human respiration}

We wish to estimate \CO2 emission from human respiration.
We give the following data
\begin{itemize}
\item human typical daily needs: 2500 kcal
\item dry pasta energy content: 350 kcal / 100g
\item massic carbon content of pasta: 30\%
\end{itemize}

\begin{enumerate}[1. ]
\item Balance the schematic respiration equation corresponding to glucose oxidation
$${\ce {C6H12O6 + 6 O2 -> 6 CO2 + 6 H2O}} + \text{energy}$$

\item How much dry pasta in grams must a human eat everyday to cover energetic needs ?

2500 / 350 $\times 100 \simeq$ 700 g.

\item What are the corresponding \CO2 emissions per person per day ?
\label{q:emissions}

700 $\times$ 0.3 $\times\frac{44}{12} \simeq $ 770 g

\item Using a rounded result (say 100g, 1 kg or 10 kg,\ldots per day depending on what you found), estimate the annual \CO2 emissions from human population ?

\textit{We take 1kg = $10^{-3}$t, then annuals emissions in Gt\CO2 are}
$ 10^{-3} \times 7.5\; 10^9\times 365 \simeq 2.7 \text{Gt} \CO2 $

\noindent Is this appreciable compared to the order of magnitude of human activities emissions ?

\textit{The typical human activities \CO2 emissions are of the order of 36 Gt\CO2 so this is appreciable.}

\item Do we have to count that in the human activities GHG emissions ? Explain.

\textit{This must not be counted because this carbon comes from food made of organic material built through photosynthesis, so taken from the air. So these emissions, though non-negligible, are compensated. If you were eating oil or coal for feeding your muscles, you would count them.}

\end{enumerate}

\subsection{Air renewal and heating power at home}

Throughout this exercise, we consider a typical square house of ground surface $S=L^2=100$m$^2$ and height $H=2.5$m, giving a volume of $V=250$ m$^3$.
We recall the perfect gas constant $R\simeq 8.3$ JK$^{-1}$mol$^{-1}$ and assume a typical temperature $T=20^{\circ}$C under pressure $P=1$ atm.

\subsubsection{Air renewal in an house}

\noindent We wish to estimate the typical volumic flow rate $D \equiv \dot{V}$ in m$^3$/hour for air renewal. 

\begin{enumerate}[1.]

\item Using the estimate of \CO2 emissions per person per day found in question \ref{q:emissions} of the previous exercise, find the \CO2 emission rate per person in kg / hour denoted by $\dot{m}_{\CO2}$.

\textit{per person:} $\dot{m}_{\CO2}$ = 1 kg\CO2 /day $\simeq$ 0.042 kg\CO2 / hour

\item We denote by $c_{\CO2}$ the \CO2 concentration in mol/L and $M_{\CO2}$ the \CO2 molar mass. Considering that 4 people are living in the house, give the relation between $D$, $c_{\CO2}$, $M_{\CO2}$ and $\dot{m}_{\CO2}$  that is satisfied in a steady state.

\textit{The emitted \CO2 must be withdrawn by air renewal so we have}
$$4\dot{m}_{\CO2} = c_{\CO2} M_{\CO2} D$$
\textit{the factor 4 coming from the number of person.}

\item We want to ensure that the \CO2 volumic concentration in the house remains below $x_{\CO2} =1000$ppm in the steady state. Infer the \CO2 concentration $c_{\CO2}$ as a function of $x_{\CO2}$ and the molar volume of air $v_{\text{air}}$.

\textit{We have} $x_{\CO2} = 10^{-3} = c_{\CO2} v_{\text{air}}$

\item Estimate the molar volume of air $v_{\text{air}}$ in L/mol. \label{q:vair}

$v_{\text{air}} = RT/P \simeq 24$ L/mol$^{-1}$

\item Conclude on the typical value of $D$.
$$D = \frac{4\dot{m}_{\CO2} v_{\text{air}}}{ x_{\CO2} M_{\CO2}} 
\simeq \frac{4\times 0.042\times 24\;10^{-3} }{10^{-3} \times 0.044} \simeq 92 \;
\text{m}^3 / h $$

\end{enumerate}

\subsubsection{Required heating power for air heating}

One reason for heating a house at temperature $T$ is to eat the incoming air at $T_{\text{ext}}$ coming from outside. We take winter conditions $T_{\text{ext}} = -5^{\circ}$C for estimates. We denote by $C_V$ the volumic specific heat of air and $C_V^m = \frac{5}{2}R$  $[\text{J\,mol}^{-1}\text{K}^{-1}]$ the molar specific heat of air.

\begin{enumerate}[1. ]
\setcounter{enumi}{5}
\item Using the estimate of question \ref{q:vair}, give the estimate for $C_V$ in JL$^{-1}$K$^{-1}$.
$$ C_V  = C_V^m / v_{\text{air}} = \frac{29}{24} \simeq 1.2 \;\text{JL}^{-1}\text{K}^{-1} \simeq 1.2 \;\text{kJm}^{-3}\text{K}^{-1}
$$

\item Give the estimate of the power $P_{\text{air}}$ required to heat up the incoming air. Is it negligible ?
$$ P_{\text{air}} = DC_V(T-T_{\text{ext}}) \simeq 92\times 1.2 \times 25 / 3600 \simeq 0.77\, kW $$
\textit{This is small but non-negligible.}

\item For double flux ventilation, say the incoming air is now at 15$^\circ$C, what is the energy gain in \% ? 

\textit{The gain is to reduce by a factor 5 / 25 so 80\%.}

\end{enumerate}

\subsubsection{Required heating power for compensating losses}

In badly isolated houses, the main reason for heating is to compensate energy losses through the envelope of the building. For simplicity, we assume that the walls and roof and in a single material, without windows (say concrete) and we neglect the losses to the ground. 

\begin{enumerate}[1. ]
\setcounter{enumi}{7}
\item What is the effective surface envelope $\Sigma$ in m$^2$ ?
$$ \Sigma = L^2 + 4LH = 200 \text{m}^2 $$

\item We denote by $\varphi$ the surfacic thermal flux through the envelope (in W/m$^2$) and $r_{\text{th}}$ its surfacic thermal resistance. 
Give the relation between $\varphi$, $r_{\text{th}}$ and the temperatures.
$$ \varphi = \frac{T-T_{\text{ext}}}{r_{\text{th}}}  $$

\item We give that the envelope width is $e=20$cm and has a thermal conductivity $\lambda = 1$ Wm$^{-1}$K$^{-1}$ (concrete). Compute $r_{\text{th}}$.
$$ r_{\text{th}} = \frac{e}{\lambda} = 0.2/1 = 0.2 \,\text{Km}^2\text{/W}$$

\item What is the estimate for $\varphi$ ? Infer the estimate of the power $P_{\text{loss}}$ required to compensate the losses. Comment.\\
\textit{We get} $\varphi \simeq 25 /0.2 = 125$ W/m$^2$. 
$P_{\text{loss}} = \varphi\Sigma = 125 \times 200 = 25$ kW

\item With rock-wool isolation with $\lambda = 0.033$ Wm$^{-1}$K$^{-1}$, quickly give the gain in energy. Compare with air renewal power.\\
\textit{This clearly divides by 30 the required power and makes it comparable to the air renewable power.}

\end{enumerate}

\subsection{The physics of the Renault Zoe}

In this exercise, we discuss in more details the energy consumption in a car.
As a basic approximation, the car dynamics along
an horizontal axis $Ox$, assuming a velocity $\vec{v} = v\;\vec{e}_x$ with $v>0$, reads
\begin{equation}
m\dv{v}{t} = F_{\text{prop}} - \mu mg - \frac{1}{2}\rho\, C_x S v^2
\end{equation}
where $F_{\text{prop}}$ is the engine mechanical force, $\mu$ is the rolling resistance coefficient corresponding to a friction force formally analogous to a Coulomb static friction force (although more complicated), and the last term is the fluid resistance of the air, or drag.\\

\noindent \textbf{Numerical data for a Renault Zoe electrical car}
\begin{itemize}
\item Mass $m=1500$ kg
\item Rolling resistance $\mu = 0.01$
\item Air mass density $\rho = 1.3$ kg/m$^3$
\item Drag coefficient $C_x = 0.35$
\item Cross sectional area $S = 2.15$ m$^2$
\item Engine efficiency $\eta = 90$\%
\item Battery capacity $Q = 52$ kWh.
\item Heat capacity of brakes : $C = 5000$ J/K
\end{itemize}

\subsubsection{Speeding up, braking and elevation}

In this section, we neglect all friction terms.

\begin{enumerate}[1.]

\item The time for a Zoe to reach $v_M=100$ km/h is $T=11.4$ s. Provide an estimate for the maximal power of the engine.\\
\textit{We use the balance of transformation from 0 to 100 km/h
$\Delta E = \frac12 m v_M^2 $ over time $T$ so that 
$P_{\text{max}} \simeq \Delta E / T \simeq 51 $ kW.
This is an underestimate since the engine has to fight friction on top of that.}

\item How speeding up and braking depend on mass ?\\
\textit{Clearly, the kinetic energy creation or destruction is directly proportional to the vehicle mass so these phases are strongly mass-dependent.}

\item During braking, the brakes are heated up. Give an estimate of the final temperature of the brakes after stopping the car from a speed of 100 km/h.\\
\textit{We use $C(T_f-T_i) = \Delta E = \frac12 m v_M^2$ with $T_i \simeq 300$ K. We obtain $T_f = T_i + m v_M^2 /2C \simeq 416$ K ($\simeq$ 140°{C}).}

\item Consider that the car climb a pass between two mountains of elevation $h=1000$ m on a $D=15$ km distance road at average speed $v=60$ km/h.
What is the typical power associated to gravity on this trip (cost in ascent, benefit in descent) ? Comment.\\
\textit{$P_{\text{gravity}} = mgh / T = mgh v / D \simeq 16.4$ kW.
This is the same order of magnitude as typical driving powers (see lectures and results below). Of course, the lower the speed, the lower the power in this configuration but the energy cost ($mgh\simeq 4.1$ kWh) remains the same. We also see that it is non-negligible if one can recover part of it through regenerative braking.}

\end{enumerate}

\subsubsection{Steady state analysis}

We consider a steady state, driving at constant speed.
Through the questions, we are going to progressively fill the Table \ref{tab:energycar}.
\begin{table}[h]
\centering
\begin{tabular}{|c|c|c|c|}
\hline
speed [km/h] &  30 & 80 & 110 \\ \hline\hline
$P_{\text{roll}}$ [kW] & 1.2 & 3.3 & 4.5 \\ \hline
$P_{\text{drag}}$ [kW] & 0.3 & 5.4 & 14.0 \\ \hline
$P_{\text{prop}}$ [kW] & 1.5 & 8.7 & 18.5 \\ \hline
$E_{\text{cons}}$ [kWh/100km] & 5.6 & 12.1 & 18.7  \\ \hline
$A$ [km] & 936 & 429 & 277  \\ \hline
$P_{\text{prop}}^{\text{wind}}$ [kW] & 2.5 & 13.8 & 27.5 \\ \hline
\end{tabular}
\caption{Typical order of magnitude for Zoe at constant speed.}
\label{tab:energycar}
\end{table}

\begin{enumerate}[1.]

\item Give the analytic expression of the power consumption $P_{\text{prop}}$ of the car. Identify two contributions $P_{\text{roll}}$ and $P_{\text{drag}}$
and fill up the corresponding lines in Table \ref{tab:energycar}. Comment.\\
\textit{For a steady state, we get $ P_{\text{prop}} = P_{\text{roll}} + P_{\text{drag}}$ with $ P_{\text{roll}} = \mu mg v$ and $P_{\text{drag}} = \frac{1}{2}\rho_, C_x Sv^3 $. The drag is almost negligible at low speeds while it is dominant at high speed.}

\item Give a simple estimate for the theoretical maximal velocity of the car ?\\
\textit{Neglecting the rolling friction at high speed and using the previous result $P_{\text{max}} \simeq 51$ kW, we get
$v_{\text{max}} = \left(\frac{2P_{\text{max}}}{\rho\,C_xS}\right)^{1/3} 
\simeq 170$ km/h.}

\item Fill the battery energy consumption $E_{\text{cons}}$ in kWh per 100km in Table \ref{tab:energycar} ?\\
\textit{We have $E_{\text{cons}} = P_{\text{prop}} T / \eta = P_{\text{prop}} D/(\eta v)$ with $D=100$ km and $\eta$ the engine efficiency.}

\item Fill the autonomy $A$ (maximal distance without charging the battery) in Table \ref{tab:energycar} ? Comment.\\
\textit{We have $A = Q /E_{\text{cons}}$. The low speed result is not very realistic because of acceleration deceleration phases that dominates at low speeds.}

\item Qualitatively discuss possible direct and indirect effects of the mass of the car on energy consumption in the cruising regime ?\\
\textit{The mass is less problematic at cruising speed in horizontal line. A direct effect is that a larger mass increases the rolling resistance. Possible indirect effects are \\
i) $S$ usually increases with the volume / mass of the car.\\
ii) $P_{\text{max}}$ increases with the mass of the car (more expensive, more sporty!)}

\item Consider a front wind of $w=20$ km/h in opposite direction of the car. After giving the new analytical formula for the car power, 
fill up the total power $P_{\text{prop}}^{\text{wind}}$ line in Table \ref{tab:energycar}.\\
\textit{The velocity $v$ of the car in the frame of the road is unchanged. 
Only the drag force is affected, according to 
$P_{\text{drag}} = \frac{1}{2}\rho_, C_x S(v+w)^2v $ since the velocity of the air relative to the car is now $v+w$. The effect is non negligible !}

\end{enumerate}

\subsection{Carbon and forest}
\label{sec:carbon}

We discuss several orders of magnitude around carbon in forest. We first give a few figures:
\begin{itemize}
\item as seen in the tutorial, a French forest managed in a renewable way typically produces 6 m$^3$/ha/year of fresh wood and typically gathers an hundred trees each containing 2.5 m$^3$ of wood.
\item dry mass of fresh (or green) wood is 65\% of the total mass (the rest being water)
\item the carbon content of dry wood is 50\% of the mass
\item the roots of a tree represent about 20\% of the unburied mass
\item there is about 40 Mkm$^2$ of forest on Earth
\end{itemize}

\begin{enumerate}[1.]
    \item What is the conversion factor between 1t of carbon and 1t of CO$_2$ ?
    
    \textit{we have $n_{\CO2} = n_{\rm C}$ so $m_{\CO2} = \frac{M_{\CO2}}{M_{\rm C}} m_{\rm C}$ with $M_{\CO2} =12+16\times 2= 44$g/mol and $M_{\rm C} = 12$g/mol so 1tC $\Rightarrow \frac{44}{12}= 3.67$ t\CO2}

    \item What is the CO$_2$ content of 1 ton of fresh wood ?

 	\textit{With fw = fresh wood and dw = dry wood:
$m_{\text{dw}} = 0.65 m_{\text{fw}}$ and $m_{\rm C} = 0.5 m_{\text{dw}}$ so $m_{\CO2} = 3.67 \times m_{\rm C} = 3.67\times 0.5 \times 0.65 m_{\text{fw}} = 1.2 m_{\text{fw}}$ (almost one to one !) depends of course a bit on the nature of the wood.}

    	\item Assuming the soil content is of the same order of magnitude in forest, how much CO$_2$ is typically stored in trees per hectare of forest ?
   
\textit{$m_{\CO2} = 2\times 1.2 \times m_{\text{fw}} = 2.4 m_{\text{fw}}$ with 2 for the soil and 1.2 for the extra 20\% from the roots. If we take the volume of fresh wood to be estimated as $V_{\text{fw}} = 100\times 2.5 = $ 250 m$^3$/ha, it corresponds to a mass of $m_{\text{fd}} = \rho_{\rm fw}V_{\rm fw}$ with $\rho_{\rm fw}$ the volumetric mass of wood. $\rho_{\rm fw}$ is not given so we have to make an educated guess. We know that most fresh wood is floating on water, almost floating, so that the volumetric mass or density should be close to 1 t/m$^3$ and we take that as an input, we are not going to make a big error. Of course, dry wood is floating much better with a lower density but something with a density probably close to 0.65 after removing the water content. 
In the end, we get the stored carbon per hectare as $m_{\CO2} = 2\times 1.2 \times m_{\text{fw}} = 2.4 \times 250 = 600 $ t\CO2/ha. }

    		\item How much CO$_2$ is released by burning 1t of dry wood ? Why do people say that it is a good idea to fight global warming by heating a house with dry wood ? Do you see a better use for wood ?
    	
    	\textit{Again, $m_{\CO2} = 3.67 \times m_{\rm C} = 3.67\times 0.5 m_{\text{dw}} = 1.84 m_{\text{dw}}$ so 1t of dry wood releases 1.84t\CO2.
\\
Usually, dry wood is seen as a renewable energy that is neutral with respect to \CO2 since the emitted \CO2 is in principle compensated by the regrowth of the wood. Of course, when you burn wood, you do emit \CO2 so all depends on what is done aside. You need to regrow the forest, which is not guaranteed and you need to wait. If you burn a tree, you have to wait 50 years for it to be adult again, so one has to be very careful in the time scales. Usually, in a renewable manner, one can extract 1 or 2 trees per hectare. Then, it is \CO2 neutral and better then burning fossil energies.\\
Yet, wood is a way to store carbon is a way probably more useful than for energy. Instead of burning the tree, one could convert it into an house or furniture or simple goods that will store carbon away from the forest and for decades. There is a real carbon compensation. Another one that is foreseen is to use wood energy in a heater equipped with \CO2 capture. Then, forest is pumping carbon out of the atmosphere (providing oxygen in passing and other benefits) and \CO2 is not emitted in the atmosphere. Yet, one needs a good storage for \CO2 at the end of the process and there is a competition for land use and biodiversity because it means industrial forest management. }

	\item Worldwide, the average ratio of sequestration of CO$_2$ for forest is estimated to be 2.7 tCO$_2$/ha/yr.
	 \begin{enumerate}
	 \item Recall the order of magnitude of the global CO$_2$ emissions by human activities. 
	 
\textit{	 In the lecture, we gave $m_{\CO2}$ = 36 Gt\CO2 / yr.}
	
	 \item Give an estimate of the surface of forest required to compensate these emissions. Comment.

\textit{The surface required to compensate is $S = m_{\CO2} / r_s$ with $r_s$ the sequestration rate. We get $S = 13.3$ Gha $= 133$ Mkm$^2$. This is more than 3 times the available size of forest of 40 Mkm$^2$. We are not going to make it only this way. First goal is to reduce.}

	 \end{enumerate}

\end{enumerate}

\subsection{Ocean elevation}

We discuss the order of magnitude of ocean elevation due to global warming. The ocean surface represents 70.8\% of the Earth surface. We recall the radius of the Earth $R=6370$km.

\begin{figure}[h]
\centering
\includegraphics[height=5cm]{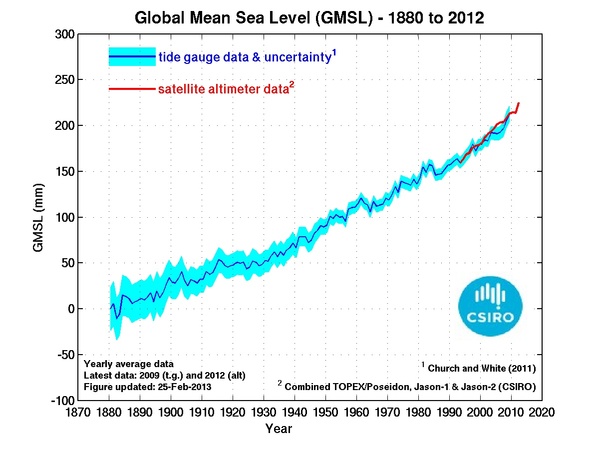}
\includegraphics[height=4cm]{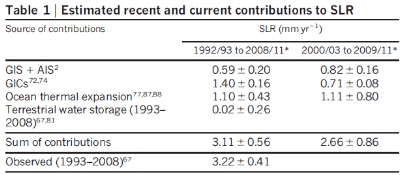}
\caption{Left: Sea level rise. Right: Contributions to the annual rise of the sea level. SLR = Sea Level Rise ; GIS = Greenland Ice Sheet ; AIS = Antarctic Ice Sheet ; GIC = Glaciers and Ice Caps (source: Hanna et al., Nature 2013)}
\label{fig:rise}
\end{figure}

\begin{enumerate}[1.]

\item We give the thermal expansion coefficient of water at 15$^{\rm o}$C $\alpha \displaystyle = \frac{1}{V}\frac{\partial V}{\partial T} \simeq 150\,10^{-6}$K$^{-1}$. Assuming that the average ocean depth that has been heated to be 1000m and after recalling the typical increase of temperature over the last century, give the estimate of the thermal expansion increase. Compare with the data of figure~\ref{fig:rise}.

\textit{Let $H = 1$ km be the typical depth that is thermally expanded of a value $\Delta H \ll H$. Considering the volume $S\times H$ of a column of surface $S$, we can write $\Delta H = \alpha H \Delta T$ with $\Delta T$ the temperature increase. Taking $\Delta T \simeq 1$ K over the last century (140 years more precisely, from 1880 to 2020), we get $\Delta H \simeq 15$ cm = 150 mm, so that thermal expansion is typically $1.1$ mm/yr, in (magical) agreement with the table.}

\begin{figure}[h]
\centering
\includegraphics[height=7cm]{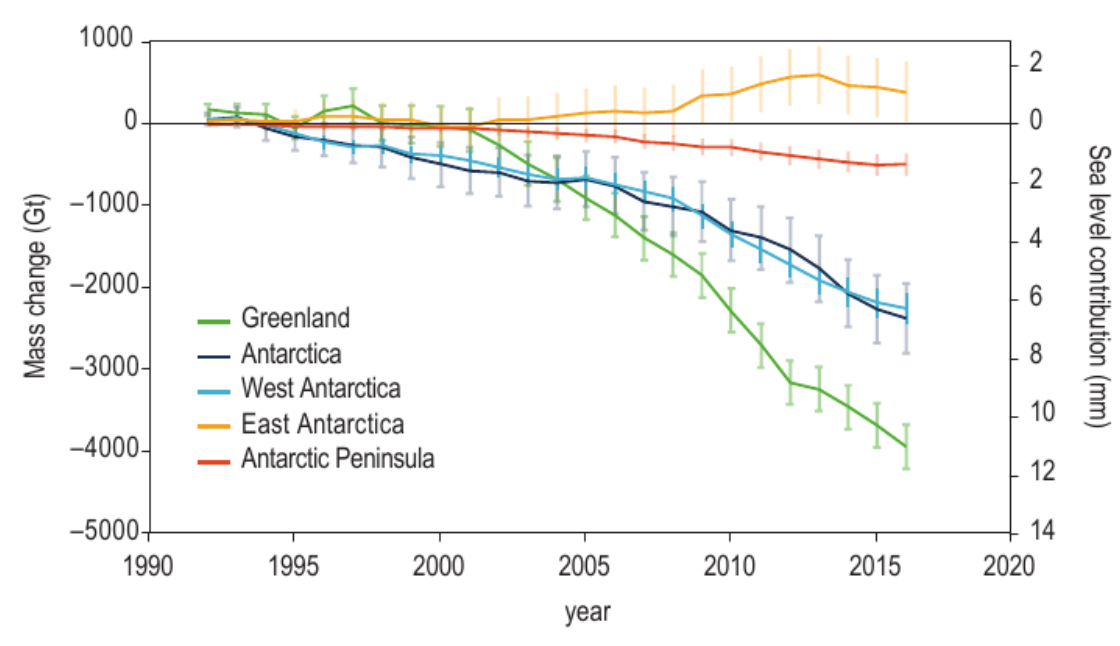}
\caption{Mass variation of continental ice and corresponding sea level increase.}
\label{fig:melting}
\end{figure}

    \item Explain  the conversion factor appearing in figure~\ref{fig:melting} to convert a mass of ice sheet in Gt to a sea level elevation in mm. Use 4000 Gt as an example.

\textit{We use mw = melted water. We have to convert the mass of ice in a volume of added water using $m_{\rm ice} = m_{\rm mw} = \rho_{\rm mw} V_{\rm mw}$ and second the volume added converts into a sea level increase $\Delta H$ simply using $V_{\rm mw} = S\Delta H$ with $S$ the surface of the Earth. The conversion factor should then be: $ \Delta H = (1/\rho_{\rm mw}S) m_{\rm ice}$. Using $\rho_{\rm mw} \simeq 1000$kg/m$^3$, $S\simeq 361$ Mkm$^2$, we get $1/\rho_{\rm mw}S \simeq 2.77\;10^{-3}$ mm/Gt, so that 4000 Gt of ice converts into 11.08 mm which is in agreement with the two $y$-axis of Figure~\ref{fig:melting}.}

\begin{figure}[h]
\centering
\includegraphics[height=6cm]{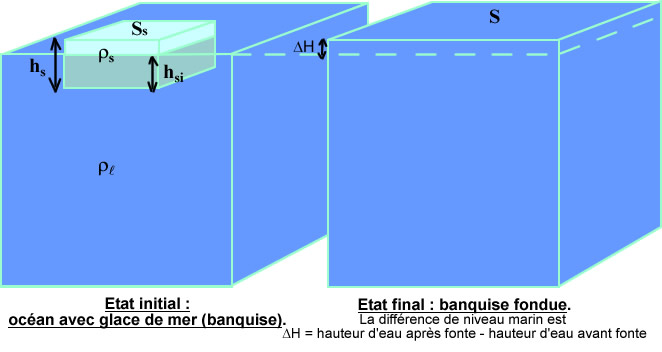}
\caption{Modelling sea ice melting.}
\label{fig:seaice}
\end{figure}

\item When sea ice melts, it could lead to an increase of sea level if the density of water coming from the sea ice (melted water) is different from the density of sea water. To estimate such increase we use the model of figure~\ref{fig:seaice} with the following notations and numbers

\centering
\begin{tabular}{|c|c|c|}
\hline
Symbol & meaning & value \\
\hline
$\rho_s$ & volumetric mass density of sea ice & 917 kg/m$^3$ \\
$S_s$ & total surface of sea ice & $34\,10^{6}$ km$^2$ \\
$h_s$ & total thickness of sea ice & 1.89 m\\
$h_{si}$ & average thickness of \textit{immersed} sea ice & \\
$\rho_f$ & volumetric mass density of \textit{melted} sea ice & 999.8 kg/m$^3$ \\
$\rho_\ell$ & volumetric mass density of sea water & 1027.68 kg/m$^3$\\
$S$ & total surface of the oceans (including sea ice) & $361\,10^{6}$ km$^2$ \\
$\Delta H$ & variation of the sea level & \\
\hline
\end{tabular}

\begin{enumerate}

\item In the initial state, find a relation between $\rho_s$, $\rho_\ell$, $h_s$ and $h_{si}$ from Archimeda's principle.

\textit{The volume of displaced water compensate for the weight of sea, which gives the mass equilibrium $\rho_\ell S_s h_{si} = \rho_s S_s h_s = m_s$ so that $h_{si} = (\rho_s/\rho_\ell) h_s$.}

\item Give the expression of the volume of water $S\Delta H$ added by the melting of sea ice as a function of $S_s$, $\rho_s$, $h_s$, $\rho_f$ and $\rho_\ell$. What happens if $\rho_f = \rho_\ell$ ?

\textit{When all the sea ice has melted, it puts a mass $m_s = \rho_s S_s h_s$ of ice converted into the same mass of melted water occupying a volume $V_f = m_s/\rho_f$, to which we must subtract the volume $h_{si} S_s = m_s/\rho_\ell$ previously occupied by immersed sea ice so that
$$ S\Delta H = m_s\left(\frac 1 {\rho_f} - \frac{1}{\rho_\ell}\right) =  \rho_s S_s h_s \left(\frac 1 {\rho_f} - \frac{1}{\rho_\ell}\right)$$
If $\rho_\ell = \rho_f$, it gives zero: we recover the usual result that with pure (unsalted) water, an ice cube melting will not change the level of water. Here, the difference is that melted sea ice is less dense that sea water that is salted, inducing a change of sea level.}

\item Give the estimated value of $\Delta H$. Comment.

\textit{Using the results and the table, we get $$ \Delta H =  h_s\rho_s \frac{S_s}{S} \left(\frac 1 {\rho_f} - \frac{1}{\rho_\ell}\right) \simeq 4.43\;\text{mm}$$
So this effect is negligible with respect to the two others.}

Inspired from 
\url{https://planet-terre.ens-lyon.fr/article/fonte-banquise-2005-10-06.xml}
\end{enumerate}

\end{enumerate}

\subsection{Temperature within a Yakhchal}

In Iran, people used to collect ice during winter and store it in dome-like structures called Yakhchals. It is a semi-buried building which walls are made of adobe and thatch. The ice occupies the lower buried section (about half of the underground volume) and is isolated from the bottom soil and from the air by thatch. The air is allowed to flow from a door carved into the side walls at ground level and the vent hole at the very top of the dome. The total height of a yakhchal is typically 10-15 m. For more than 3 months, in summer, the temperatures in Iran are above 30$^{\rm o}$C. In the following, you are expected to model the problem as realistically as possible. Briefly explain your assumptions, if you make any.

\noindent \textbf{Question: Determine the effective thermal resistance of the dome-like structure walls.}

\noindent \textbf{Answer:} If you assume a wall made of 0.5 m of adobe and 0.2 m of thatch and a Yakhchal radius of 5 m, one gets : $R_{th}=\sum \frac{e}{\lambda S}=\frac{1}{2\pi R^2}\left(\frac{0.5}{0.6}+\frac{0.2}{0.1}\right) = 0.018$ K/W.

\noindent \textbf{Question:  On a schematic drawing, determine what are the thermal exchange mechanisms at play.}

\noindent \textbf{Answer:} There is the conduction through the side walls, conduction from the ground to the ice through thatch, conduction of air through the vent hole. There are the radiations from the soil, the walls and the top vent hole. Convection from the door to the vent hole also has to be considered.

\noindent \textbf{Question: What are the predominant mechanisms in your opinion and why?}

\noindent \textbf{Answer:} The radiation is dominated by the one due to the vent hole (because of the external temperature radiating through it). The conduction by air is negligible. The thermal resistance of the thatch layer (taken to be 0.1 m) between the ground temperature assumed to be of about 15$^{\rm o}$C is of 0.005 K/W and dominates that of the walls. Convection has to be considered.

\noindent \textbf{Question: If the ice is warmed only through its contact with the soil and if the ice storage is full at the beginning of summer, how much is lost during these 3 months?}

\noindent \textbf{Answer:} In this question, we simplify the problem by considering only the conduction from the ground to the ice through the hatch. This is not entirely true, but makes the calculation easier. Then, assuming the ground is at 15$^{\rm o}$C and the ice at 0$^{\rm o}$C, the power due to this thermal exchange is : $P=\iint \overrightarrow{\varphi}.\overrightarrow{dS}= \frac{\lambda S}{e}\left(0-15\right)=2'600$ W. During 3 months, this produces about 2$\times 10^7$ kJ corresponding to 60'500 kg of melted ice. Initially, we have about 120 m$^3$ of ice, i.e. 110 tons of ice. There therefore is about 56$\%$ of the ice that is melted at the end of summer.

\noindent Numerical data that may be of use:
\begin{itemize}
    \item[$\bullet$] Thermal conductivity: Adobe 0.6 W/m.K, Thatch 0.1 W/m.K, Soil 0.3 W/m.K, Ice 1.7 W/m.K, Air 0.025 W/m.K.
    \item[$\bullet$] Density: Adobe 1'700 kg/m$^3$, Thatch 250 kg/m$^3$, Soil 1'000 kg/m$^3$, Ice 920 kg/m$^3$, Air 1.2 kg/m$^3$.
    \item[$\bullet$] Specific heat capacity: Adobe 900 J/kg.K, Thatch 1'800 J/kg.K, Soil 1'900 J/kg.K, Ice 2'108 J/kg.K, Air 29'000 J/kg.K.
    \item[$\bullet$] Latent heat of melting: Water/Ice 334 kJ/kg
    \item[$\bullet$] The Stefan-Boltzmann constant $\sigma=5.68\times 10^{-8}$ W.K$^{-4}$.m$^{-2}$
    \item[$\bullet$] Convective heat transfer for air: $h=10$ W.m$^{-2}$.Km$^{-1}$
\end{itemize}

\subsection{The Dubai Frame}

The Dubai Frame is a new (2018) building in Dubai (United Arab Emirates). The golden frame contains (golden) PV panels. 1'200 m$^2$ of amorphous silicon photovoltaic glass panes were installed on the facade, with a transparency of 20\%.

\noindent\textbf{Question: Estimate the maximum power that can be produced by these PV panels.} 

\noindent\textbf{Answer:} $P_{max}=1200$ m$^2\times 34$ Wp/m$^2 = 40.8$ kW.

\noindent\textbf{Question: Estimate the office surface for which this solar power can provide the energy for.}

\noindent\textbf{Answer:} Assuming offices work for 8 hours per day, the average power needed for an office is : $P_{office}\simeq\frac{250 \, \rm kWh.m^{-2}.yr^{-1}}{365 \rm \, days \times 8 \, hours}\simeq 85$ W/m$^2$. PV panels can therefore provide the power needed for about 480 m$^2$ of office.

\noindent\textbf{Question: Estimate the energy that can be produced daily, and compare it to the energy brought by sunlight during the same period.}

\noindent\textbf{Answer:} Assuming the PV panels are lit 10 hours per day, $P_{max}$ typically corresponds to an energy of: $E_{max}=P_{max}\times 10$ h $\simeq 408$ kWh/day.\\
    The surface occupied by the building is typically $A=100\times 10 = 1000$ m$^2$. On this surface, the Sun brings about $E_{sun}=5.9$ kWh/m$^2$.day $\times$ 1000 m$^2$  = 5.9 MWh/day.\\
    The PV panels are thus able to harvest about 7\% of the incoming solar energy.

\noindent\textbf{Question: Quantitatively discuss the use of these 20\%-transparency PV modules with respect to other existing PV modules.}

\noindent\textbf{Answer:}
\begin{itemize}
        \item[$\bullet$] Amorphous silicon PV panels are less efficient than crystalline silicon PV panels, and even more so when they are rendered transparent. If, on the same surface, crystalline silicon PV panels had been used, the harvested energy could have been multiplied by almost a factor 8.
        \item[$\bullet$] However, this transparency also allows sunlight to penetrate within the building and may help saving lighting energy.
        \item[$\bullet$] Moreover, these golden PV panels have a lesser visual impact than crystalline PV panels, much to the satisfaction of architects.
\end{itemize}

\noindent\textbf{Question: Discuss the pros and cons of the use of PV panels in this situation.}

\noindent\textbf{Answer:}
 \begin{itemize}
        \item \textbf{Pros}
            \begin{itemize}
                \item[$\bullet$] In Dubai, the GHI is important, so that it would have been a shame not to take advantage of this technology.
                \item[$\bullet$] The use of solar energy for powering an office allows the energy to be used immediately, without it needing to be stored.
                \item[$\bullet$] The Dubai Frame is a flagship building, and having PV panels used for the construction may serve as an incentive for other buildings to use solar energy.
            \end{itemize}
        \item \textbf{Cons}
            \begin{itemize}
                \item[$\bullet$] PV panels fabrication involve not-so-clean chemistry, so that, to fully assess the impact of these PV panels, one would need to run a cradle-to-cradle life cycle analysis.
                \item[$\bullet$] The PV panels used here are not optimized: their production capability is much less than that of crystalline silicon PV panels, and, due to the shape of the building itself, they are not at the optimal orientation (facing South and tilted by about 15$\rm ^o$). Their efficiency will therefore not be maximum. One may also expect the neighboring buildings may cast some shadow onto the panels, preventing them to produce electricity, at least during some hours in the day.
                \item[$\bullet$] More PV panels could have been used, for instance on the rooftop. These could have been crystalline silicon PV panels and may have been more efficient.
                \item[$\bullet$] These PV panels can, at most, provide the energy for about 500 m$^2$. Considering their orientation and real efficiency, the actual energy production is anecdotal for the building. One could then question the cost and carbon footprint of this installation.
            \end{itemize}
    \end{itemize}
\noindent Numerical data that may be of use:
\begin{itemize}
    \item[$\bullet$] Global Horizontal Irradiation (GHI) in Dubai: 5.9 kWh/m$^2$ daily.
    \item[$\bullet$] Average energy consumption of office buildings: 250 kWh/m$^2$.yr.
    \item[$\bullet$] Data on PV panels:
\end{itemize}

\begin{table}[!h]
\begin{ruledtabular}
\begin{tabular}{llrr}
&&Visible light transparency & Peak power (Wp/m$^2$)\\
\hline
& Low Transparency & 10\% & 40\\
Amorphous Si PV & Medium Transparency & E0\% & 34\\
& High Transparency & 30\% & 28\\
\hline
&&&\\
Crystalline Si PV & High Density & 0\% & 241\\
& Low Density & 15\% & 64\\
\end{tabular}
\end{ruledtabular}
\end{table}

\subsection{Insulation and energy consumption}

This fall and winter, there have been talks about power shutdowns in case the energy demand is larger than the energy consumption. Although dwellings are usually not the main source of energy consumption, people have been asked to reduce their use of heating to safeguard the energy network. In this exercise, we would like to compare the energy consumption of a well-insulated house with that of a badly-insulated house.

\begin{figure}[h]
\centering
\includegraphics[width=0.6\textwidth]{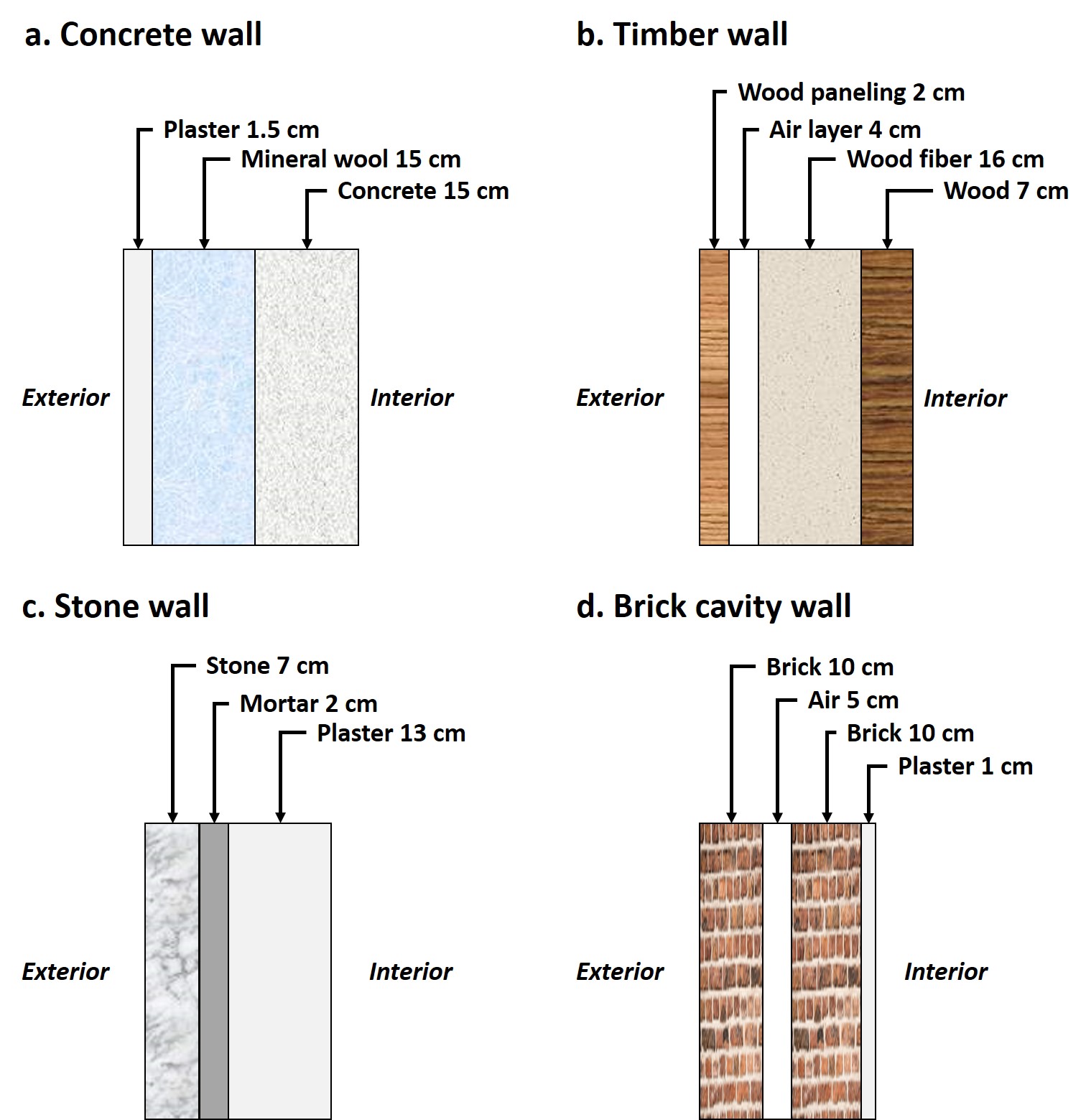}
\caption{A few typical wall compositions: a. Concrete wall, b. Timber (wood) wall, c. Stone wall, d. Brick cavity wall.}
\label{fig:Wall_compo}
\end{figure}

\begin{figure}[h]
\centering
    \includegraphics[width=0.6\textwidth]{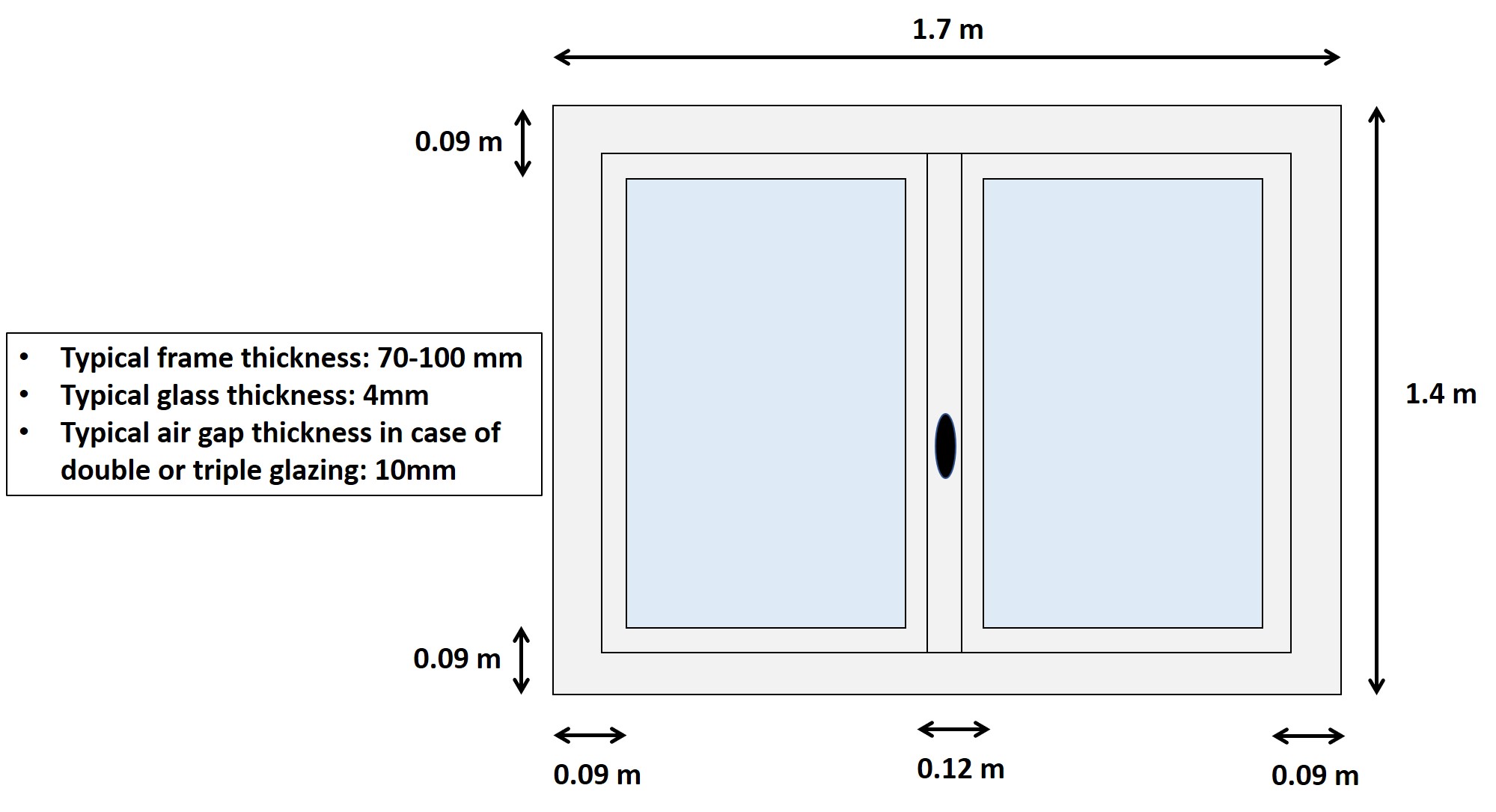}
    \caption{Typical window size.}
    \label{fig:Window_compo}
\end{figure}

We will consider a single storey house during a winter day.

\begin{itemize}
    \item[$\bullet$] Please detail the assumptions you make, such as the surface of the house, its location, the external temperature, etc.
    \item[$\bullet$] For simplicity, you do not need to distinguish between walls, roof and floor, and you can assume that they all have the same composition. You can either choose \textbf{ONE} of the typical wall compositions shown in figure \ref{fig:Wall_compo}, or propose your own wall composition.
    \item[$\bullet$] The dimensions of a typical window are given in figure \ref{fig:Window_compo}. You can however modify those \textbf{IF} you explicitly say so.
    \item[$\bullet$] Keep your house design simple!!
\end{itemize}

\begin{enumerate}
    \item \textbf{Preliminary calculation \#1 -} Determine the equivalent thermal resistance of 1 m$^2$ of the wall composition you have chosen.

\noindent\textbf{Answer:} For the concrete wall of surface $S=1$ m$^2$, the thermal resistance depends on the thermal conductivities $\lambda_i$ and thicknesses $e_i$:
\begin{eqnarray}
R_{wall}&=&\frac{1}{S}\sum \frac{e_i}{\lambda_i}\\
&=&4.64 \,\text{K.W}^{-1}?
\end{eqnarray}
    
    \item \textbf{Preliminary calculation \#2 -} Determine the equivalent thermal resistance of the window you have chosen if it is a single-pane window (simple vitrage).

\noindent\textbf{Answer:} Let us consider a window of dimensions shown in figure \ref{fig:Window_compo}, with a plastic frame. The total surface of the window is 2.38 m$^2$, of which 1.708 m$^2$ is glass, and 0.672 $m^2$ is plastic. For a single-pane window, the thermal resistance is:
\begin{eqnarray}
    R_{window}&=&\left(\frac{\lambda_{plastic}S_{plastic}}{e_{plastic}}+\frac{\lambda_{glass}S_{glass}}{e_{glass}}\right)^{-1}\\
    &=&3\times 10^{-3}\,\text{K.W}^{-1}.
\end{eqnarray}
    
    \item \textbf{Preliminary calculation \#3 -} Determine the equivalent thermal resistance of the window you have chosen if it is a double-glazed window. 

\noindent\textbf{Answer:} For the same window with double-glazing (thicknesses given in figure \ref{fig:Window_compo}, the thermal resistance of the glass can be obtained by considering 3 resistances in series (glass, air, glass):
\begin{eqnarray}
    R_{glass}&=&\frac{1}{1.708}\left(2\frac{0.004}{0.8}+\frac{0.01}{0.025}\right)\\
    &=&0.24 \,\text{K.W}^{-1}.
\end{eqnarray}
For the entire window, one obtains $R_{window}=0.18\,\text{K.W}^{-1}$.
    
    \item Determine the energy consumption (the heating energy that is necessary) to maintain the inside temperature at  $T_{int}=19^{\rm o}$C for 1 hour, in the following cases:
    \begin{enumerate}
        \item \textbf{Case 1 - A box -} Consider a house without any openings, i.e. a box, which walls have your chosen wall composition. 

\noindent\textbf{Answer:} In the following, we will consider a 1-storey house of surface 10$\times$10 m$^2$, and 3 m in height. We will assume that the outside temperature is constant during the hour, and equal to 0$^{\rm o}$C. 

\noindent In this question, we will neglect convection and radiation, and only consider heat transfer through the walls, the roof and the ground. We considered that the ground was at an average temperature of 15$^{\rm o}$C (mean value between the maximum and minimum values throughout the year). The power dissipated by conduction is:
\begin{eqnarray}
    P_{box}&=&\left(\frac{100+4\times 30}{R_{wall}}\right)\left(0-19\right)+\left(\frac{100}{R_{wall}}\right)\left(15-19\right)\\
    &=&986\,\text{W}    
\end{eqnarray}
So the energy necessary to maintain the inside temperature during 1 hour is $E_{box}$=986 Wh.
        
        \item \textbf{Case 2 - A single-pane house -} Consider a house with single pane windows (please explicit the number of windows and their dimensions). Any comments?

\noindent\textbf{Answer:} Let use consider a house with 10 identical windows, for a total window surface of 23.8 m$^2$. The power dissipated by conduction is:
\begin{eqnarray}
    P_{single\,window}&=&\left(\frac{220-23.8}{R_{wall}}+\frac{10}{R_{window}}\right)\left(0-19\right)+\left(\frac{100}{R_{wall}}\right)\left(15-19\right)\\
    &=&64\,\text{kW}    
\end{eqnarray}
So the energy necessary to maintain the inside temperature during 1 hour is $E_{single\,window}$=64 kWh. This makes a huge difference, and is mainly due to the very poor insulating properties of glass. To improve this, one should use double- or triple-glazed windows.
        
        \item \textbf{Case 3 - A double-glazed house -} Consider a house with double-glazed windows (please explicit the number of windows and their dimensions). Any comments?

\noindent\textbf{Answer:} By the same calculation, one obtains an energy necessary to maintain the inside temperature during 1 hour is $E_{double\,glazed}$=1.9 kWh. This makes a huge difference, and is mainly due to the very good insulating properties of air.
        
        \item \textbf{Case 4 - A badly-insulated house -} Consider the influence of air drafts for instance. Any comments?

\noindent\textbf{Answer:} Let us model the heat loss through air drafts by a 5 mm gap on the length of all windows (either the top or the bottom). This is pretty large, but gives an order of magnitude of the effect (a 2 mm gap on each length is not uncommon). The corresponding surface is $0.005\times 1.7\times 10 = 0.085$ m$^2$. Then the power dissipated through convection is:
\begin{eqnarray}
    P_{draft}&=&hA(T_s-T_\infty)\\
    &=&h\times 19\times 0.085 
\end{eqnarray}
For a wind speed of 5 m.s$^{-1}$ (almost no wind), one obtains an additional energy $E_{wind,\,5}=160$ Wh, whereas for a wind speed of 30 m.s$^{-1}$ (storm), the energy required to compensate the temperature loss through the draft plummets at $E_{wind,\,30}=500$ Wh. This becomes non-negligible compared to the energy calculated in Case 3.
    \end{enumerate}
\end{enumerate}

\textit{Numerical data that may be of use:}
\begin{itemize}
    \item[$\bullet$] Thermal conductivity: Air 0.025 W/m.K, Brick 0.8 W/m.K, Concrete 0.5 W/m.K, Glass 0.8 W/m.K, Mineral wool 0.035 W/m.K, Mortar 1 W/m.K, Plaster 0.25 W/m.K, Plastic 0.2 W/m.K, Stone 2 W/m.K, Wood 0.15 W/m.K, Wood fiber 0.04 W/m.K, Wood paneling 0.15 W/m.K
    \item[$\bullet$] Convective heat transfer for air : $h=100$ W.m$^{-2}$.Km$^{-1}$ for a wind speed of 5 m.s$^{-1}$
    \item[$\bullet$] Convective heat transfer for air : $h=300$ W.m$^{-2}$.Km$^{-1}$ for a wind speed of 30 m.s$^{-1}$
\end{itemize}


\section{List of useful materials}
\label{sec:List of useful materials}

\begin{table}[!h]
\begin{ruledtabular}
\begin{tabular}{lr}
Cooking thermometer & 5-20 \$ \\
\hline
Arduino-based O$_2$ sensor & 50-70 \$ \\
\hline
Arduino-based CO$_2$ sensor & 100 \$ \\
\hline
Arduino-based humidity sensor & 5-10 \$ \\
\hline
FLIR IR camera (for smartphone) & 200-350 \$ \\
\hline
Lux meter & 30-60 \$ \\
\hline
Wind speed sensor & 30-100 \$\\
\hline
Laser meter & 10-20 \$\\
\hline
Pt100 thermal sensor to solder & 3-10 \$\\
\end{tabular}
\end{ruledtabular}
\caption{\label{tab:table1}%
List of useful materials and their typical price unit as of fall 2022. }
\end{table}

\end{document}